\documentclass[10pt,pra,aps,groupedaddress,twocolumn,nofootinbib,floatfix,raggedbottom]{revtex4-2}
\usepackage{graphicx}
\usepackage{amsmath}
\usepackage{amssymb}
\usepackage{siunitx}
\usepackage{xcolor}
\usepackage{braket}
\usepackage{titlesec}
\usepackage{hyperref}
\hypersetup{
    pdfborder={0 0 0}
}
\usepackage{orcidlink}

\titlespacing{\section}{0pt}{*0.6}{*0.3}
\titlespacing{\section}{0pt}{0.4\baselineskip}{0.3\baselineskip}
\titlespacing{\subsection}{0pt}{0.4\baselineskip}{0.3\baselineskip}
\titlespacing{\subsubsection}{0pt}{0.4\baselineskip}{0.3\baselineskip}

\begin{document}

\title{On a local hidden-variable model for postselected EPR correlations}

\author{Daniele Cenni\orcidlink{0000-0002-9761-9111}}
\affiliation{SIAF, University of Florence, 50141 Firenze, Italy}
\date{\today}

\begin{abstract}
A theoretical and numerical analysis of a Bell-type experiment is
presented, where linear polarization measurements on nonmaximally
entangled two-qubit states are modeled through a sigmoid-based threshold
mechanism that violates the fair-sampling assumption required by the CHSH
inequality. Numerical simulations characterize how closely this local
construction can reproduce quantum-like correlations in the postselected
detected sample, for both the $\ket{\Phi_r}$ and $\ket{\Psi_r}$ states
over a range of entanglement parameters. For the tested
configurations, the model produces CHSH-type violations only in the
postselected detected sample through a setting-dependent detection
mechanism, operating outside the assumptions of both the CHSH inequality
and Mermin's $82.8\%$ efficiency bound. The model is not in conflict with loophole-free Bell experiments, which specifically close the detection loophole exploited here.
The CH--Eberhard analysis in the one-detector-per-side configuration yields
$J < 0$ throughout the full range of entanglement parameters, as
required for a strictly local model.
The model is further extended with a
local adaptive-threshold mechanism implementing one-sided memory.
A paired comparison shows that this specific memory rule produces a
systematic and reproducible shift in the CH--Eberhard parameter $J$
relative to the fixed-threshold model, while leaving the
CH--Eberhard analysis qualitatively unchanged under the tested
conditions. The results characterize how a local, setting-dependent
postselection mechanism shapes Bell-test statistics across a range of
entanglement parameters, within the specific configurations and threshold
rules tested.
\end{abstract}

\maketitle

\section{Introduction}
The Einstein--Podolsky--Rosen paradox~\cite{Einstein1935} and Bell's theorem \cite{Bell1964} established a fundamental incompatibility between locality and the completeness of quantum mechanical description. Bell's work demonstrated that no local hidden-variable theory could account for all quantum mechanical predictions. After that, the experimental verification of Bell's inequality has evolved through several stages.

Freedman and Clauser's experiment in 1972~\cite{Freedman1972} provided the first experimental evidence supporting quantum mechanics over local hidden-variable theories.

Aspect et al. in the early 1980s~\cite{Aspect1981, Aspect1982experimental, Aspect1982} addressed the locality loophole through time-varying analyzers that changed orientation during the photons' flight, ensuring spacelike separation between measurements.

Zeilinger et al. fully enforced for the first time the condition of locality~\cite{Weihs1998}, and later extended the work over $144$ kilometers \cite{Ursin2007}, establishing the persistence of quantum correlations over macroscopic distances.

Other experiments closed the detection loophole~\cite{Christensen2013, Giustina2013}, which could be exploited to build a local hidden-variable model, and the major locality and detection loopholes were simultaneously closed by Hensen et al.~\cite{Hensen2015}, Giustina et al.~\cite{Giustina2015}, and Shalm et al.~\cite{Shalm2015} under the stated experimental assumptions.

Local models exploiting the detection loophole have shown that setting-dependent detection efficiency can produce apparent Bell violations in the postselected sample while remaining strictly local; the appropriate criterion in this context is the CH (or CH--Eberhard) inequality, which does not assume fair sampling.

The present work introduces a local hidden-variable model in which the measurement process for nonmaximally entangled photon pairs is modeled through a scoring system combined with a sigmoid-based threshold mechanism. This construction violates the fair-sampling assumption required by the CHSH inequality (hence the CHSH inference does not apply) through a setting-dependent detection mechanism, and produces CHSH-type violations only in the postselected detected sample of photons. The CH--Eberhard analysis yields $J < 0$ throughout the full range of entanglement parameters: the model is strictly local by construction and is not in conflict with loophole-free Bell experiments, which specifically close the detection loophole exploited here. A model variant incorporates an adaptive-threshold mechanism to explore memory-driven effects and their influence on the CH--Eberhard inequality, confirming the same conclusion under locally history-dependent detection. These results provide a basis for examining how specific detection-level mechanisms---both static and history-dependent---shape Bell-test outcomes within a fully local framework.

\section{Related Work}

Werner~\cite{Werner1989} showed that a state of a composite quantum system is classically correlated if it can be approximated by convex combinations of product states, and Einstein--Podolsky--Rosen correlated otherwise. Barrett~\cite{Barrett2002} extended this to generalized measurements.

Since entanglement and Bell nonlocality are inequivalent~\cite{Augusiak2015, Bowles2016}, determining whether a given state is actually local requires tools that go beyond entanglement tests alone. This motivates the construction of explicit LHV models capable of approximating quantum-like statistics over a continuum of measurement settings, which contributes to assessing locality beyond the scope of finite-setting Bell tests.

Local hidden-variable models exploiting the detection loophole were studied systematically starting with Pearle~\cite{Pearle1970}, who showed that data rejection in a local model can produce apparent Bell violations. In Pearle's model the joint detection rate depends on the relative angle between the two settings.

Branciard~\cite{Branciard2011} showed that the set of postselected local correlations forms a polytope strictly larger than the standard local polytope, and that in the CHSH scenario its non-trivial facets are of CH--Eberhard type; Bell inequalities of this form should therefore be tested rather than the
standard CHSH bound $S \leq 2$. His analysis assumes setting-independent detection efficiency.

Larsson~\cite{Larsson2014} notes that fair sampling is not equivalent to constant detector efficiency: local realist models with constant but lowered efficiency that reproduce quantum-mechanical predictions exist, while the conditional correlations after detection give a violation but the underlying predictions do not violate the CHSH inequality.

Gisin and Gisin~\cite{GisinGisin1999} gave an explicit analytical construction for the singlet state, in which the detection probability at the first station is proportional to $|\lambda\cdot a|$, where $\lambda$ is a hidden variable with uniform a priori probability distribution and $a$ is the local measurement direction; this reproduces the quantum correlation exactly with an average detection efficiency of $75\%$, and produces apparent CHSH violations in the postselected detected sample. At least one detector always fires, and the model does not account for nonmaximally entangled states.

The present model covers the nonmaximally entangled states $|\Phi_r\rangle$ and $|\Psi_r\rangle$ for $r \in [-1,1]$, allowing the dependence of detection efficiency and approximation accuracy on entanglement strength to be characterized systematically. It approximates rather than exactly reproduces quantum correlations, with approximation quality quantified by the RMSE between simulated and quantum correlations as a function of $r$. The detection probabilities at the two stations are independent by construction (in the Gisin--Gisin model they are not), and the no-click outcome is explicitly modeled. The hidden-variable-dependent detection, with setting-dependent efficiency for $|r|<1$, constitutes the operative loophole mechanism of the model.

\section{Theoretical Framework}\label{sec:theoretical_framework}

A source of entangled photons is considered in one of the nonmaximally entangled states $\mathcal{B}$
\begin{align}
\ket{\Phi_r} &= \frac{1}{\sqrt{1+r^2}}(\ket{HH} + r \ket{VV}) \label{eq:phi_plus} \\
\ket{\Psi_r} &= \frac{1}{\sqrt{1+r^2}}(\ket{HV} + r \ket{VH}) \label{eq:psi_plus}
\end{align}
where $H$ and $V$ denote horizontal and vertical polarization states respectively, and $r$ is the entanglement parameter. For $r=\pm1$, quantum states are maximally entangled. Hence

\begin{align}
\ket{\Phi_r} =
\begin{cases}
\ket{\Phi_r^+}, & \text{if } r > 0 \\
\ket{\Phi_r^-}, & \text{if } r < 0
\end{cases}
\end{align}

\begin{align}
\ket{\Psi_r} =
\begin{cases}
\ket{\Psi_r^+}, & \text{if } r > 0 \\
\ket{\Psi_r^-}, & \text{if } r < 0
\end{cases}
\end{align}

Polarization measurements are performed on spacelike separated entangled photons. The measurement basis for linear polarization at angle $\phi$ is
\begin{align}
\ket{\phi} &= \cos(\phi)\ket{H} + \sin(\phi)\ket{V} \\
\ket{\phi^{\perp}} &= -\sin(\phi)\ket{H} + \cos(\phi)\ket{V}
\end{align}
where $\ket{H}$ and $\ket{V}$ are the horizontal and vertical polarization states, respectively.

\begin{figure}[htbp]
\centering
\includegraphics[width=\columnwidth]{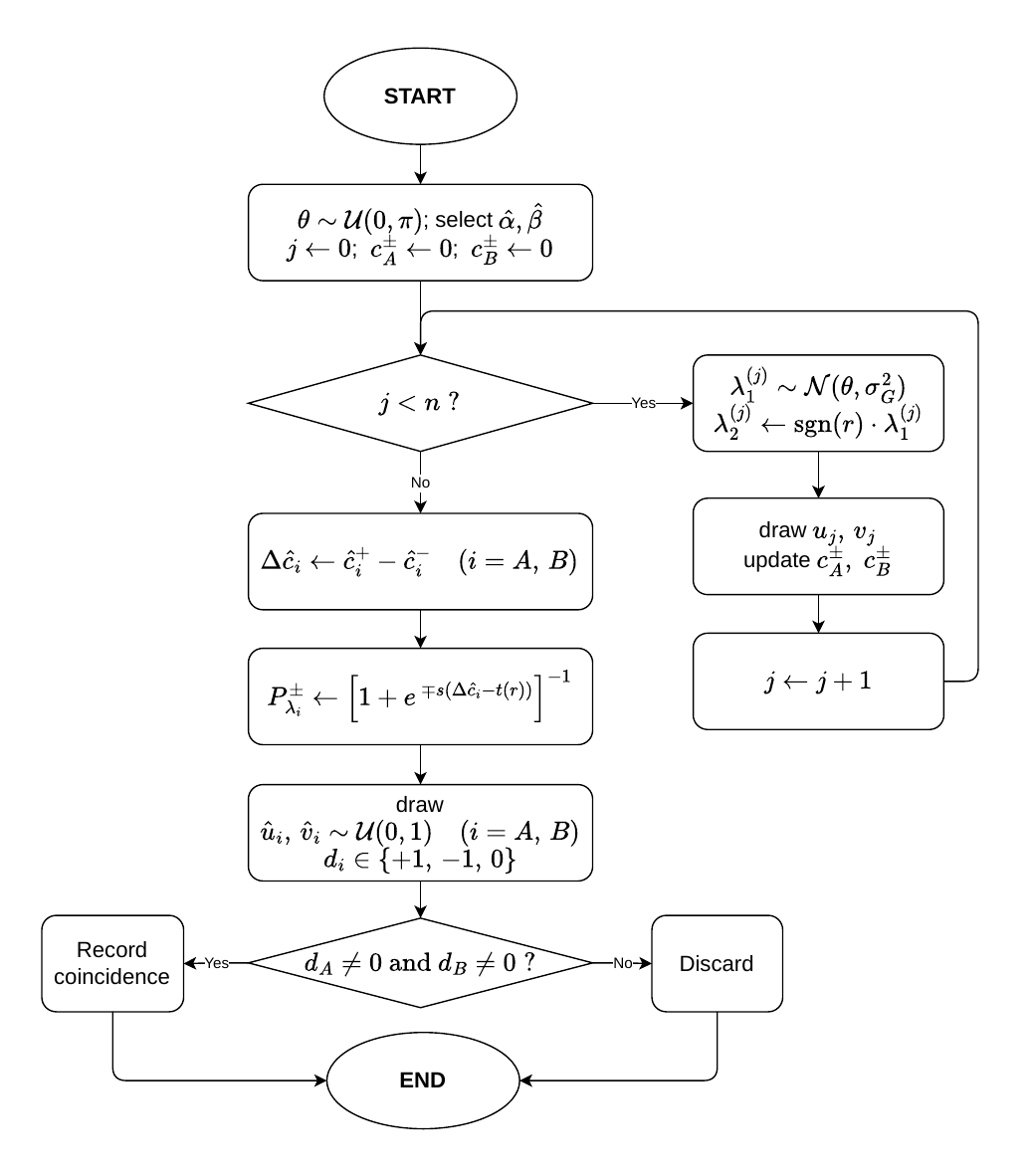}
\caption{Flowchart of the sigmoid LHV model: from hidden-variable components $\lambda^{(i)}$ to cumulative scores $c_i^\pm$, net score $\Delta\hat{c}$, sigmoid detection probabilities $P_{\lambda_i}^\pm$, and outcome $d_{\lambda_i} \in \{+1,-1,0\}$.}
\label{fig:flowchart}
\end{figure}

The polarization measurement observable at angle $\phi$ is constructed from the projectors $\hat{\sigma}_\phi = \ket{\phi}\bra{\phi} - \ket{\phi^\perp}\bra{\phi^\perp}$:
\begin{equation}
\hat{\sigma}_\phi =
\begin{pmatrix}
\cos^2\phi - \sin^2\phi & 2\sin\phi\cos\phi \\
2\sin\phi\cos\phi & -(\cos^2\phi - \sin^2\phi)
\end{pmatrix}
\end{equation}
Applying the double-angle identities $\cos(2\phi) = \cos^2\phi - \sin^2\phi$ and $\sin(2\phi) = 2\sin\phi\cos\phi$ gives $\hat{\sigma}_\phi = \cos(2\phi)\hat{\sigma}_z + \sin(2\phi)\hat{\sigma}_x$.
The quantum correlation $E_{\Phi_r}(\hat{\alpha},\hat{\beta},r) =
\bra{\Phi_r}\hat{\sigma}_{\hat{\alpha}}\otimes
\hat{\sigma}_{\hat{\beta}}\ket{\Phi_r}$ for $\ket{\Phi_r} = (\ket{HH} + r\ket{VV})/ \sqrt{1+r^2}$ is then
\begin{align}
E_{\Phi_r}(\hat{\alpha},\hat{\beta},r) &=
\frac{1}{1+r^2}\bigl[
\bra{HH}\hat{\sigma}_{\hat{\alpha}}\otimes
\hat{\sigma}_{\hat{\beta}}\ket{HH} \nonumber\\
&\quad + r^2\bra{VV}\hat{\sigma}_{\hat{\alpha}}
\otimes\hat{\sigma}_{\hat{\beta}}\ket{VV}
\nonumber\\
&\quad + r\bra{HH}\hat{\sigma}_{\hat{\alpha}}
\otimes\hat{\sigma}_{\hat{\beta}}\ket{VV}
\nonumber\\
&\quad + r\bra{VV}\hat{\sigma}_{\hat{\alpha}}
\otimes\hat{\sigma}_{\hat{\beta}}
\ket{HH}\bigr]
\end{align}
Evaluating each term using $\hat{\sigma}_\phi\ket{H} = \cos(2\phi)\ket{H} + \sin(2\phi)\ket{V}$ and $\hat{\sigma}_\phi\ket{V} = \sin(2\phi)\ket{H} - \cos(2\phi)\ket{V}$
\begin{align}
\bra{HH}\hat{\sigma}_{\hat{\alpha}}\otimes
\hat{\sigma}_{\hat{\beta}}\ket{HH} &=
\cos(2\hat{\alpha})\cos(2\hat{\beta})\\
\bra{VV}\hat{\sigma}_{\hat{\alpha}}\otimes
\hat{\sigma}_{\hat{\beta}}\ket{VV} &=
\cos(2\hat{\alpha})\cos(2\hat{\beta})\\
\bra{HH}\hat{\sigma}_{\hat{\alpha}}\otimes
\hat{\sigma}_{\hat{\beta}}\ket{VV} &=
\sin(2\hat{\alpha})\sin(2\hat{\beta})\\
\bra{VV}\hat{\sigma}_{\hat{\alpha}}\otimes
\hat{\sigma}_{\hat{\beta}}\ket{HH} &=
\sin(2\hat{\alpha})\sin(2\hat{\beta})
\end{align}
Collecting the terms:
\begin{equation}
E_{\Phi_r}(\hat{\alpha}, \hat{\beta}, r) = \frac{(1+r^2)\cos(2\hat{\alpha})
\cos(2\hat{\beta}) +
2r\sin(2\hat{\alpha})\sin(2\hat{\beta})}{1+r^2}
\end{equation}
and similarly for $\ket{\Psi_r}$.

Therefore, the quantum correlations for $\ket{\Phi_r}$ and $\ket{\Psi_r}$ predicted by quantum mechanics are
\begin{align}
E_{\Phi_r}(\hat{\alpha}, \hat{\beta}, r) &= \cos(2\hat{\alpha})\cos(2\hat{\beta})+\frac{2r}{1+r^2}\sin(2\hat{\alpha})\sin(2\hat{\beta}) \label{eq:QMcorr_Phi} \\
E_{\Psi_r}(\hat{\alpha}, \hat{\beta}, r) &= -\cos(2\hat{\alpha})\cos(2\hat{\beta})+\frac{2r}{1+r^2}\sin(2\hat{\alpha})\sin(2\hat{\beta}) \label{eq:QMcorr_Psi}
\end{align}
where $\hat{\alpha} \in \{a, a'\}$ and $\hat{\beta} \in \{b, b'\}$ are the linear polarization angles of the first and second station respectively.

The state of the first photon $\lambda_1=(\theta_1,...,\theta_n)$ is randomly sampled from a Gaussian distribution $\mathcal{N}(\theta, \sigma_G^2)$ with mean $\theta$ and standard deviation $\sigma_G=\lvert (1-r^2) \rvert / (\sqrt{3}(1 + r^2))$, and that of the second photon $\lambda_2$ is defined as
\begin{align}
\lambda_2 =
\begin{cases}
\lambda_1, &\text{ if } \mathcal{B} \in \{\ket{\Phi_r^+}, \ket{\Psi_r^-}\} \\
-\lambda_1, &\text{ if } \mathcal{B} \in \{\ket{\Phi_r^-}, \ket{\Psi_r^+}\}
\end{cases}
\end{align}

The sign of $\lambda_2$ encodes the correlation structure of the state. The polarization angle $\theta$ is randomly drawn from a uniform distribution $\mathcal{U}(0, \pi)$, with a probability density function $p(x)=1/\pi$ if $x \in [0, \pi)$, and $0$ elsewhere.

The modified sign function $\hat{\operatorname{sgn}}(r)$ and the modified Heaviside function $\hat{\mathcal{H}}$ are defined as

\begin{align}
\hat{\operatorname{sgn}}(r) &=
\begin{cases}
+1 & \text{if } r \geq 0 \\
-1 & \text{if } r < 0
\end{cases} \label{eq:sign}
\end{align}

\begin{align}
\hat{\mathcal{H}}(x) &=
\begin{cases}
1 & \text{if } x > 0 \\
0 & \text{if } x \leq 0
\end{cases} \label{eq:heaviside}
\end{align}

Given the orientations of the polarizers $\hat{\alpha}$ and $\hat{\beta}$, the cumulative scores $(c_1^{+}, c_1^{-})$ and $(c_2^{+}, c_2^{-})$ accumulate weighted contributions from $n$ hidden-variable components for the first and second photon respectively, where the Heaviside function $\hat{\mathcal{H}}$ selects at each component whether the contribution is added to the $+$ or $-$ branch

\begin{align}
c_1^+ &= \sum_{i=1}^{n} \hat{\mathcal{H}}\left(\frac{1 + \sigma_{\mathcal{B}}E_{\mathcal{B}}(\hat{\alpha}, \lambda_1^{(i)}, r)}{2} - u_i\right) \times \nonumber\\
&\times \left[ w_{a}(\hat{\alpha}, \lambda_1^{(i)}, r) \cos^2(\hat{\alpha} - \hat{\operatorname{sgn}}(r)\lambda_1^{(i)}) + \right. \nonumber\\
&\quad\left. w_{b}(\hat{\alpha}, \lambda_1^{(i)}, r) \sin^2(\hat{\alpha} - \hat{\operatorname{sgn}}(r)\lambda_1^{(i)}) \right] \label{eq:score1_plus}
\end{align}

\begin{align}
c_1^- &= \sum_{i=1}^{n} \hat{\mathcal{H}}\left(u_i - \frac{1 + \sigma_{\mathcal{B}}E_{\mathcal{B}}(\hat{\alpha}, \lambda_1^{(i)}, r)}{2}\right) \times \nonumber\\
&\quad \times \left[ w_{c}(\hat{\alpha}, \lambda_1^{(i)}, r) \cos^2(\hat{\alpha} - \hat{\operatorname{sgn}}(r)\lambda_1^{(i)}) + \right. \nonumber\\
&\quad \quad \left. w_{d}(\hat{\alpha}, \lambda_1^{(i)}, r) \sin^2(\hat{\alpha} - \hat{\operatorname{sgn}}(r)\lambda_1^{(i)}) \right] \label{eq:score1_minus}
\end{align}

\begin{align}
c_2^+ &= \sum_{i=1}^{n} \hat{\mathcal{H}}\left(\frac{1 + \sigma_{\mathcal{B}}E_{\mathcal{B}}(\hat{\beta}, \lambda_2^{(i)}, r)}{2} - v_i\right) \times \nonumber\\
&\quad \times \left[ w_{a}(\hat{\beta}, \lambda_2^{(i)}, r) \cos^2(\hat{\beta} - \hat{\operatorname{sgn}}(r)\lambda_2^{(i)}) + \right. \nonumber\\
&\quad \quad \left. w_{b}(\hat{\beta}, \lambda_2^{(i)}, r) \sin^2(\hat{\beta} - \hat{\operatorname{sgn}}(r)\lambda_2^{(i)}) \right] \label{eq:score2_plus}
\end{align}

\begin{align}
c_2^- &= \sum_{i=1}^{n} \hat{\mathcal{H}}\left(v_i - \frac{1 + \sigma_{\mathcal{B}}E_{\mathcal{B}}(\hat{\beta}, \lambda_2^{(i)}, r)}{2}\right) \times \nonumber\\
&\quad \times \left[ w_{c}(\hat{\beta}, \lambda_2^{(i)}, r) \cos^2(\hat{\beta} - \hat{\operatorname{sgn}}(r)\lambda_2^{(i)}) + \right. \nonumber\\
&\quad \quad \left. w_{d}(\hat{\beta}, \lambda_2^{(i)}, r) \sin^2(\hat{\beta} - \hat{\operatorname{sgn}}(r)\lambda_2^{(i)}) \right] \label{eq:score2_minus}
\end{align}

\noindent where $\lambda_1^{(i)}$ and $\lambda_2^{(i)}$ are the $i$-th components of $\lambda_1$ and $\lambda_2$ respectively, $u_i$ and $v_i$ are independent random numbers sampled from a uniform distribution $\mathcal{U}(0,1)$, $\sigma_{\mathcal{B}} = \{1 \text{ if } \mathcal{B}=\ket{\Phi_r}, -1 \text{ if } \mathcal{B}=\ket{\Psi_r}\}$.

The hidden-variable components $\lambda_1^{(i)}$ are defined in the same angular space $[0,\pi)$ as the polarizer angles $\hat{\alpha}$. The expressions $\cos^2(\hat{\alpha} - \hat{\operatorname{sgn}}(r)\lambda_1^{(i)})$ and $\sin^2(\hat{\alpha} - \hat{\operatorname{sgn}}(r)\lambda_1^{(i)})$ represent squared projections between the polarizer direction and the local hidden-variable direction (and similarly for $\lambda_2^{(i)}$ and $\hat{\beta}$).

The modified sign function $\hat{\operatorname{sgn}}(r)$ selects the angular projection direction in Eqs.~(\ref{eq:score1_plus})--(\ref{eq:score2_minus}): for $r > 0$ the hidden-variable direction enters unchanged, while for $r < 0$ it is reflected, consistently with the anticorrelated structure of $\lambda_2 = -\lambda_1$ for $|\Phi_r^-\rangle$ and $|\Psi_r^+\rangle$.

The modified Heaviside $\hat{\mathcal{H}}$ implements a probabilistic test on the uniform draw: the component contributes to the $+$ branch when $u_i < [1+\sigma_{\mathcal{B}}E_{\mathcal{B}}(\hat{\alpha},\lambda_1^{(i)},r)]/2$ and to the $-$ branch otherwise, and analogously for $v_i$ at the second station. Thus each hidden-variable component contributes to exactly one branch of the cumulative score.

The weights $w(\rho, \lambda, r)$ in Eqs.~(\ref{eq:score1_plus})--(\ref{eq:score2_minus}) are defined as

\begin{align}
w_{a}(\rho, \lambda, r) &= \frac{P_{++}(\rho, \lambda, r)}{P_{++}(\rho, \lambda, r) + P_{--}(\rho,\lambda, r)}\label{eq_score_weights_a}\\
w_{b}(\rho, \lambda, r) &= \frac{P_{--}(\rho, \lambda, r)}{P_{++}(\rho, \lambda, r) + P_{--}(\rho, \lambda, r)}\label{eq_score_weights_b}\\
w_{c}(\rho, \lambda, r) &= \frac{P_{+-}(\rho, \lambda, r)}{P_{+-}(\rho, \lambda, r) + P_{-+}(\rho, \lambda, r)}\label{eq_score_weights_c}\\
w_{d}(\rho, \lambda, r) &= \frac{P_{-+}(\rho, \lambda, r)}{P_{+-}(\rho, \lambda, r) + P_{-+}(\rho, \lambda, r)}\label{eq_score_weights_d}
\end{align}
where $P_{ij}(\rho, \lambda, r)$ is the quantum mechanical coincidence probability for the state
$\ket{\Phi_r}$, evaluated using only the local polarizer angle $\rho \in
\{\hat{\alpha}, \hat{\beta}\}$ and the local hidden-variable component $\lambda \in \{\lambda_1^{(i)}, \lambda_2^{(i)}\}$ as its two angular arguments. No information from the remote station enters this evaluation.

The quantum correlation $E_\mathcal{B}(\rho, \lambda^{(i)}, r)$ in Eqs.~(\ref{eq:score1_plus})--(\ref{eq:score2_minus}) is evaluated with the local polarizer angle $\rho$ and the local hidden-variable component $\lambda^{(i)}$ as its two angular arguments, where $\lambda^{(i)}$ takes the place of the remote station's polarizer angle. The detection probability at each station therefore depends on the local measurement setting without nonlocal information, and this is the mechanism through which fair sampling is violated.

The weights $w_a$, $w_b$, $w_c$, $w_d$ are normalized ratios of local quantum-like coincidence probabilities $P_{ij}(\rho,\lambda^{(i)},r)$, evaluated using only the local polarizer angle $\rho$ and the local hidden-variable component $\lambda^{(i)}$; no information from the remote station enters this evaluation.

The scoring mechanism incorporates the quantum mechanical structure of the state through locally evaluated coincidence probabilities, without inter-station communication at any step.

The denominators in Eqs.~(\ref{eq_score_weights_a})--(\ref{eq_score_weights_d}) are non-zero whenever the corresponding branch of Eqs.~(\ref{eq:score1_plus})--(\ref{eq:score2_minus}) is entered. This follows from the identity $P_{++}+P_{--} = (1+E_\mathcal{B})/2$, which implies that the branch using $P_{++}+P_{--}$ is entered only when $(1+E_\mathcal{B})/2 > 0$; the complementary branch, entered when $u_i \geq [1+\sigma_{\mathcal{B}}E_{\mathcal{B}}]/2$ with $u_i < 1$, requires $P_{+-}+P_{-+} = (1-E_\mathcal{B})/2 > 0$.

The coincidence probabilities $P_{ij}(\rho,\lambda,r)$ are the quantum mechanical joint detection probabilities for the state $\ket{\Phi_r}$, obtained from the Born rule $P_{ij} = \bra{\Phi_r}\hat{\Pi}_i(\rho)\otimes\hat{\Pi}_j(\lambda)\ket{\Phi_r}$, where $\hat{\Pi}_+(\theta) = \ket{\theta}\bra{\theta}$ and $\hat{\Pi}_-(\theta) = \ket{\theta^\perp}\bra{\theta^\perp}$ are the projectors onto the measurement basis states.

Applying $\hat{\Pi}_+(\rho)\otimes\hat{\Pi}_+(\lambda)$ to $\ket{\Phi_r}$ gives
\begin{align}
&\bigl(\hat{\Pi}_+(\rho)\otimes
\hat{\Pi}_+(\lambda)\bigr)\ket{\Phi_r} =
\nonumber\\
&\quad= \frac{\cos(\rho)\cos(\lambda)+r\sin(\rho)\sin(\lambda)}{\sqrt{1+r^2}}
\ket{\rho}\ket{\lambda}
\end{align}
and therefore the four coincidence probabilities are

\begin{align}
P_{++}(\rho, \lambda, r) &= \frac{1}{1+r^2} \left(\cos(\rho)\cos(\lambda) + r\sin(\rho)\sin(\lambda)\right)^2 \label{eq:prob_pp} \\
P_{--}(\rho, \lambda, r) &= \frac{1}{1+r^2} \left(\sin(\rho)\sin(\lambda) + r\cos(\rho)\cos(\lambda)\right)^2 \label{eq:prob_mm} \\
P_{+-}(\rho, \lambda, r) &= \frac{1}{1+r^2} \left(-\cos(\rho)\sin(\lambda) + r\sin(\rho)\cos(\lambda)\right)^2 \label{eq:prob_pm} \\
P_{-+}(\rho, \lambda, r) &= \frac{1}{1+r^2} \left(\sin(\rho)\cos(\lambda) - r\cos(\rho)\sin(\lambda)\right)^2 \label{eq:prob_mp}
\end{align}

The detection probabilities $P_{\lambda_i}^+$ and $P_{\lambda_i}^-$ for the '+' and '-' channel respectively are defined with sigmoid functions
\begin{equation}
P_{\lambda_i}^+ = \frac{1}{1 + \exp\!\bigl[-s\,(\Delta \hat{c}_i - t(r))\bigr]} \label{eq:detection_plus}
\end{equation}
\begin{equation}
P_{\lambda_i}^- = \frac{1}{1 + \exp\!\bigl[-s\,(-\Delta \hat{c}_i - t(r))\bigr]} \label{eq:detection_minus}
\end{equation}
where $s$ is the steepness, $\Delta \hat{c}_1 = (\hat{c}_1^+ - \hat{c}_1^-)$ and $\Delta \hat{c}_2 = \sigma_{\mathcal{B}}(\hat{c}_2^+ - \hat{c}_2^-)$ are the net scores of the first and second photon, respectively, while $\hat{c}_1^+$ and $\hat{c}_1^-$ are the normalized cumulative scores

\begin{align}
\hat{c}_1^+ &= \frac{c_1^+}{c_1^+ + c_1^-} &
\hat{c}_1^- &= \frac{c_1^-}{c_1^+ + c_1^-}\label{eq:net_scores_1} \\
\hat{c}_2^+ &= \frac{c_2^+}{c_2^+ + c_2^-} &
\hat{c}_2^- &= \frac{c_2^-}{c_2^+ + c_2^-}\label{eq:net_scores_2}
\end{align}

The scores $\Delta \hat{c}_1$, $\Delta \hat{c}_2$ (derived from Eqs.~\ref{eq:score1_plus}--\ref{eq:score2_minus}) determine, through Eqs.~(\ref{eq:detection_plus})--(\ref{eq:detection_minus}), the probabilities of detection in the $+$ and $-$ channels.

The correlation sign factor $\sigma_{\mathcal{B}}$ selects between correlated and anticorrelated states by flipping the sign of the second photon's net score in the sigmoid argument, ensuring that the detection probabilities at station~2 reflect the correlation structure of the state.

The sigmoid functions in Eqs.~(\ref{eq:detection_plus})--(\ref{eq:detection_minus}) map net scores to detection probabilities in $[0,1]$, provide a smooth threshold transition controlled by the steepness $s$, and their symmetric structure ensures that the two detection channels are treated equivalently.

The threshold function $t(r)$ is defined as

\begin{equation}
    t(r) = \frac{\tau_0 + \tau_1 r^2 + \tau_2 r^4 + \tau_3 r^6}{1 + \tau_4 r^2 + \tau_5 r^4},
    \label{eq:threshold}
\end{equation}

This rational function is symmetric in $r$, bounded and smooth over the full
range of entanglement parameters, and exhibits a maximum near $|r|=1$; its parameters are determined by cross-validation, as described in the following section.

For the state $\ket{\Psi_r}$, the same expressions as in Eqs.~(\ref{eq:prob_pp})--(\ref{eq:prob_mp}) are used, since the anticorrelation is implemented through sign flip of the second photon's net score in Eq.~(\ref{eq:detection_minus}).

For each photon $\lambda_i$, random numbers $\hat{u}_i, \hat{v}_i \sim \mathcal{U}(0,1)$ are sampled to determine the measurement outcome

\begin{equation}\label{eq:measurement_outcome}
d_{\lambda_i} =
\begin{cases}
+1, & \text{if } P_{\lambda_i}^+ > \hat{u}_i \text{ and } P_{\lambda_i}^- \leq \hat{v}_i \\
-1, & \text{if } P_{\lambda_i}^+ \leq \hat{u}_i \text{ and } P_{\lambda_i}^- > \hat{v}_i \\
0, & \text{otherwise}
\end{cases}
\end{equation}
where $d_{\lambda_i} = +1$ indicates detection in the $+$ channel, $d_{\lambda_i} = -1$ indicates detection in the $-$ channel, and $d_{\lambda_i} = 0$ indicates no detection, while $\hat{u}_i$ and $\hat{v}_i$ model the noise resulting from experimental imperfections.

The $d_{\lambda_i} = 0$ outcome covers two physically distinct cases: neither channel fires (no-click) and both channels fire simultaneously (double-click).

Coincidences are recorded by comparing the detection outcomes. The single detection efficiencies $\eta_1$ and $\eta_2$ are determined by the mechanism described above. The correlation function $E_{\mathcal{B}}(\hat{\alpha},\hat{\beta})$ for polarizer orientations $\hat{\alpha}$ and $\hat{\beta}$ is calculated from the coincidence statistics

\begin{equation}
E_{\mathcal{B}}(\hat{\alpha},\hat{\beta}) = \frac{N_{++} + N_{--} - N_{+-} - N_{-+}}{N_{++} + N_{--} + N_{+-} + N_{-+}}
\end{equation}
where $N_{ij}$ represents the number of coincidences with outcome $i$ at the first analyzer and $j$ at the second analyzer. The value of the Bell parameter function $S$ is calculated using the CHSH inequality \cite{Clauser1969}

\begin{equation}
S = E_{\mathcal{B}}(a,b) - E_{\mathcal{B}}(a,b') + E_{\mathcal{B}}(a',b) + E_{\mathcal{B}}(a',b')
\end{equation}

Figure~\ref{fig:flowchart} shows a flowchart of the model. Starting  from the hidden variable $\theta$ and the $n$ Gaussian components $\lambda^{(i)}$, the cumulative scores $c_i^\pm$ are accumulated via the weighted Heaviside mechanism and normalized to yield the net score $\Delta\hat{c}$, which determines the sigmoid detection  probabilities $P_{\lambda_i}^\pm$ and the outcome $d_{\lambda_i} \in \{+1,-1,0\}$. Each station operates independently on local information only; a coincidence is recorded when both stations yield $d_{\lambda_i} \neq 0$, constituting the postselection that violates fair sampling.

The detection probability depends on the measurement settings $\{\hat{\alpha}, \hat{\beta}\}$ through the correlation functions $E_{\mathcal{B}}(\hat{\alpha}, \lambda_1, r)$ and $E_{\mathcal{B}}(\hat{\beta}, \lambda_2^{(i)}, r)$ in Eqs.~(\ref{eq:score1_plus})--(\ref{eq:score2_minus}). Since the random numbers $u_i$ and $v_i$ filter different subsets of hidden-variable components depending on the measurement angles, the average detection efficiency $\langle\eta(\hat{\alpha})\rangle$ (or $\langle\eta(\hat{\beta})\rangle$) is setting-dependent for $|r|<1$
\begin{equation}
\langle\eta(\hat{\alpha})\rangle = \mathbb{E}_{\lambda, U}[P(\lambda, \hat{\alpha}, U)]
\end{equation}
where $\mathbb{E}_{\lambda, U}$ denotes the expectation value over hidden variables $\lambda$ and the random samples $U = \{u_i\}_{i=1}^n$ or $\{v_i\}_{i=1}^n$ introduced in Eqs.~(\ref{eq:score1_plus})--(\ref{eq:score2_minus}), and $P(\lambda, \hat{\alpha}, U)$ is the detection probability defined in Eqs.~(\ref{eq:detection_plus})--(\ref{eq:detection_minus}).

This hidden-variable-dependent postselection, with setting-dependent efficiency for $|r|<1$, constitutes the operative detection loophole and violates the fair-sampling assumption required by the CHSH inequality, unlike the case of uniform efficiency loss independent of measurement settings.

\section{Model's Parameters}
The Gaussian distribution $\mathcal{N}(\theta, \sigma_G^2)$ is used to model the hidden-variable states. The standard deviation $\sigma_G(r) = |1-r^2|/(\sqrt{3}(1+r^2))$ determines the dispersion of the hidden-variable distribution around the source angle $\theta$, which decreases as $|r| \to 1$, reaching zero for maximally entangled states. It quantifies the degree of polarization uncertainty introduced by the hidden variable $\lambda$.

The parameter $n$ specifies the number of hidden-variable components independently sampled from $\mathcal{N}(\theta,\sigma_G^2)$ and contributing to the cumulative score; it directly affects the computational cost per trial.

The asymmetry of the nonmaximally entangled states $\ket{\Phi_r}$ and $\ket{\Psi_r}$ is quantified by $|1-r^2|/(1+r^2)$, which parametrizes the imbalance between the two components of their superposition.

The factor $1/\sqrt{3}$ ensures that the hidden-variable distribution $\mathcal{N}(\theta, \sigma_G^2)$ produces the same statistical dispersion as uniform noise with imbalance $|1-r^2|/(1+r^2)$, since $\text{Var}[\mathcal{N}(0,\sigma^2)] = \text{Var}[\mathcal{U}(-a,a)]$ when $\sigma = a/\sqrt{3}$.

The score weights $w(\rho,\lambda^{(i)},r)$ in Eqs.~(\ref{eq_score_weights_a})--(\ref{eq_score_weights_d}) are defined as normalized weights that modulate each hidden-variable component's contribution as a function of the relative orientation between $\lambda^{(i)}$ and the polarizer angle $\rho$. The linear combination $w_a \cos^2(\alpha-\lambda^{(i)}) + w_b \sin^2(\alpha-\lambda^{(i)})$ in Eq.~(\ref{eq:score1_plus}) (and analogously for $w_c$, $w_d$, $\beta$) provides a two-axis angular decomposition, assigning different weights to the projections along the orthogonal detector directions.

The coincidence probabilities $P(\rho, \lambda, r)$ in Eqs.~(\ref{eq:prob_pp})--(\ref{eq:prob_mp}) provide the weights that modulate each component's contribution to the cumulative score, modeling polarization-dependent detection efficiency. The use of quantum-like coincidence probabilities in the weights $w_a$--$w_d$ reflects the central question of this work: whether quantum mechanical correlation patterns, evaluated locally using only the hidden variable and the local polarizer angle, can generate a setting-dependent postselection that approximates quantum predictions in the detected sample.

The sigmoid steepness $s$ determines the sharpness of the detection probability transition: for $P = 0.10$ and $P = 0.90$, the net score satisfies $\Delta\hat{c}_i - t(r) = \pm\ln(9)/s$, so the transition from $10\%$ to $90\%$ detection probability occurs over a net-score interval of $\pm\ln(9)/s$ around the threshold.

The threshold function $t(r)$ parametrizes detection sensitivity, affecting the overall detection rate without depending on the polarizer angles. Its rational form is symmetric in $r$ and exhibits a maximum near $|r|=1$, remaining bounded and smooth for all $r$. A higher threshold requires a larger net-score imbalance $|\Delta\hat{c}_i|$ for detection, compensating for the narrower hidden-variable distribution at high $|r|$ where $\sigma_G(r)\to 0$ and net scores are more sharply peaked.

The threshold satisfies $t(r)=t(-r)$ because the detection sensitivity calibrates to the entanglement strength $|r|$: the states $|\Phi_r^+\rangle$ and $|\Phi_r^-\rangle$ (and similarly $|\Psi_r^+\rangle$ and $|\Psi_r^-\rangle$) are related by $r\to -r$ and differ only in the sign of their correlations, which is handled separately by $\sigma_{\mathcal{B}}$ in the sigmoid detection probabilities.

For $|r| \ll 1$, $t(r)$ varies approximately quadratically about the baseline value $\tau_0$
\begin{equation}
t(r) \approx \tau_0 + (\tau_1 - \tau_0\tau_4)r^2 + \mathcal{O}(r^4)
\end{equation}

The parameters $\tau_1$, $\tau_2$, and $\tau_3$ control the rate and curvature of the threshold variation with $|r|$; $\tau_4$ and $\tau_5$ control the shape of its approach to the maximum near $|r|=1$. All these parameters are determined by cross-validation and have no physical meaning beyond fitting the calibrated threshold profile.

This mechanism models polarization-dependent photodetection by accumulating orientation-dependent signal contributions before evaluating the detection outcome. The sigmoid-based detection probabilities $P_{\lambda_i}^+$ and $P_{\lambda_i}^-$ in Eqs.~(\ref{eq:detection_plus})--(\ref{eq:detection_minus}) model the competition between the two orthogonal measurement channels, where $\Delta \hat{c}_1 = (\hat{c}_1^+ - \hat{c}_1^-)$ and $\Delta \hat{c}_2 = \sigma_{\mathcal{B}}(\hat{c}_2^+ - \hat{c}_2^-)$ represent the local discrimination signals for the first and second photon, respectively. All parameters described above act locally, modulating the detection behavior at each station, and do not encode joint or nonlocal dependencies.

The model is not intended to be a full physical description of the detection process, but as an effective framework for examining detection mechanisms and for exploring the conditions under which they approximate quantum correlations.

\section{Threshold Calibration}\label{sec:calibration}

The sigmoid detection mechanism introduced in Sec.~\ref{sec:theoretical_framework} depends on a
threshold function $t(r)$ that sets the sensitivity of both detectors as a
function of the entanglement parameter $r$. Rather than fixing $t$ to an
arbitrary constant, $t(r)$ is determined empirically through a calibration
procedure that targets a specific detection efficiency $\eta_0$ for
both detectors simultaneously. The value of $\eta_0 = 0.8425$ was chosen so that the Bell parameter $S$ approaches $2\sqrt{2}$ for the polarizers' configuration that maximizes the Tsirelson bound when $r=1$. The shape of $t(r)$ is calibrated without reference to quantum correlations; the scalar $\eta_0$ encodes a quantum-referenced choice of operating point.

The calibration is performed for $r \in \{0.0, 0.1, 0.2, \ldots, 1.0\}$,
which covers all distinct entanglement regimes from the product state ($r = 0$)
to the maximally entangled Bell state $\ket{\Phi^+}$ ($r = 1$). Negative values
of $r$ are not considered because the threshold function is even by construction,
$t(r) = t(-r)$.

For each candidate threshold $t'$, the detection efficiencies $\eta_1(t')$ and
$\eta_2(t')$ of the two detectors are estimated by Monte Carlo simulation over
the current set of polarizer configurations. The threshold is chosen to minimize
\begin{align}
    F(t') &= |\eta_1(t') - \eta_0| + |\eta_2(t') - \eta_0| \nonumber \\
          &\quad + p_1 \cdot \max\!\bigl(0,\,|\eta_1(t') - \eta_2(t')| - p_2\bigr),
    \label{eq:objective_revised}
\end{align}
where $p_1 = 2.0$ controls the slope of the imbalance penalty and $p_2 = 0.024$
is the tolerance below which efficiency imbalances are left unpenalized. The
first two terms drive both detectors toward $\eta_0$; the third term suppresses
asymmetries between the two detection rates.

Threshold calibration is performed via leave-one-out cross-validation (LOO-CV)
over randomly generated polarizer configurations, ensuring that the calibrated
threshold generalizes to angle combinations not seen during the calibration itself. Twenty
independent iterations are performed for each value of $r$. In each iteration,
seven polarizer configurations are drawn at random, with all four angles $(a, a',
b, b')$ sampled independently from $[0, \pi)$. Within each iteration,
seven-fold LOO-CV is applied: six configurations form the training set and the
seventh is reserved for validation, cycling through all seven train/test
partitions and yielding seven independent threshold estimates per iteration.

For each training fold, the threshold is selected by grid search over
$N_{\mathrm{grid}} = 80$ equally spaced points in $[0.001, 0.800]$. At each
grid point $t'$, the objective $F(t')$ is evaluated by Monte Carlo simulation of
$N_{\mathrm{OPT}} = 5000$ entangled photon pairs per configuration, averaged
over $R_{\mathrm{OPT}} = 5$ independent runs. The efficiency $\eta_k(t')$ for
detector $k$ is computed as the mean over the four setting combinations $(a_i,
b_j)$ of the per-combination detection rate. The optimal threshold from the
training fold is then evaluated on the held-out configuration by simulating
$N_{\mathrm{TEST}} = 8000$ photon pairs, averaged over $R_{\mathrm{TEST}} = 10$
independent runs. An iteration is accepted if at least five of the seven folds
converge to a valid threshold (i.e., $t \in (0.001, 0.999)$ and $\eta_k > 0.1$
for both detectors). The mean and standard deviation of the accepted thresholds
across all valid iterations are recorded for each $r$ value.

The calibration yields a table of mean threshold values $\bar{t}(r)$ and their
standard deviations $\sigma_t(r)$ at the eleven points $r \in \{0.0, 0.1,
\ldots, 1.0\}$, which are fitted to the threshold function of Eq.~\ref{eq:threshold}.

The fit is performed in two steps. The eleven calibration points are first
interpolated by a shape-preserving piecewise cubic Hermite (PCHIP) spline, which
passes through all data points by construction and avoids the oscillatory
artifacts of higher-degree polynomial interpolation. The threshold function is then fitted to the spline evaluated on a fine grid of 500 points in $[0, 1]$ by nonlinear least squares with 50 random
restarts to avoid local minima, with the denominator constrained to remain
positive on $[0, 1.01]$ throughout the optimization. The two-step procedure
separates interpolation (handled exactly by the spline) from compression into a
compact analytical form (handled by the rational fit), making the fit quality
independent of the spacing of the calibration points.

\begin{figure}[htbp]
\centering
\includegraphics[width=0.85\columnwidth]{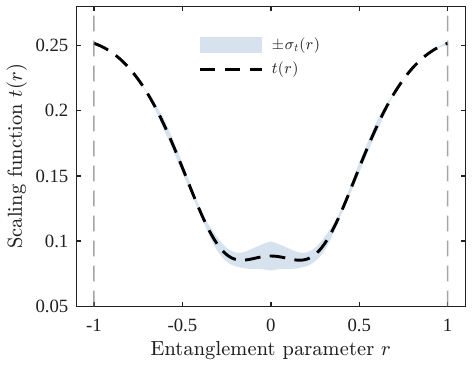}
\caption{Threshold function $t(r)$ (Eq.~\ref{eq:threshold}) fitted by cross-validation. The shaded band shows $\pm\sigma_t(r)$, the per-point standard deviation of the threshold estimates across validation folds (mean $0.0038$ across $r$). Dashed vertical lines mark $r = \pm 1$ (maximally entangled states).}
\label{fig:threshold}
\end{figure}

The fitted coefficients are $\tau_0 = 0.08850099$, $\tau_1 = 0.10348506$, $\tau_2 = 5.90902599$, $\tau_3
= -0.73932695$, $\tau_4 = 4.01967323$, $\tau_5 = 16.29316662$, with $R^2 = 0.9996$, RMSE $= 1.2\times10^{-3}$,
and maximum absolute residual $2.9\times10^{-3}$ at $r = 0$. The fitted curve
and the calibration data points with their standard deviations are shown in
Fig.~\ref{fig:threshold}.

The LOO-CV succeeded for all 11 values of $r$, with all 20 iterations accepted
at each point. The cross-validation standard deviation is $\sigma_t = 0.0038 \pm 0.0035$
(mean $\pm$ standard deviation across $r$ values), decreasing overall
from $0.011$ at $r=0$ to below $1.5\times10^{-3}$ for $|r|\geq 0.7$.
Detection efficiencies on held-out configurations achieved
$\eta_1 = 0.845 \pm 0.003$ and $\eta_2 = 0.843 \pm 0.002$ (mean $\pm$ standard
deviation across $r$ values). The $S$ values recorded in the calibration output are not informative about the model's Bell-inequality behavior, since they are computed at randomly generated
polarizer configurations not optimized for CHSH violation. The results reported in
Sec.~\ref{sec:results} are entirely independent of the calibration $S$ values.

\subsection{Reproducibility}
Each calibration run uses deterministic pseudorandom number generators
seeded from the iteration index and the value of $r$: a PCG32 generator
for the polarizer angle configurations and an XORWOW generator for the
photon-pair simulations. Because the seeds depend only on these logical
indices, and not on execution order or the degree of parallelism, the
calibration returns identical thresholds across runs on a fixed hardware and
software stack.

The raw calibration output and the fitting routine are provided as Supplementary
Material; applying the routine to the output reproduces the coefficients
$\tau_0,\ldots,\tau_5$ and the goodness-of-fit statistics reported above.

\section{Simulation Setup}

In the first setup the source that generates photon pairs in the nonmaximally entangled state is $\ket{\Phi_r}$ (maximally entangled for $r= \pm 1$). The configuration used is $\alpha=(\SI{0}{\degree}, \SI{45}{\degree})$, $\beta=(\SI{22.5}{\degree}, \SI{67.5}{\degree})$, $n=21$, $s=20$, with the threshold function $t(r)$ calibrated as described in Sec.~\ref{sec:calibration}. Relevant metrics include $S$ and its standard error of the mean; the single detection efficiencies $\eta_1$ and $\eta_2$ and their standard errors, averaged over the four measurement-setting combinations
$(\hat{\alpha},\hat{\beta})$; the Root Mean Square Error (RMSE) between the model's correlations
$\hat{E}_{\Phi_r}$ and those predicted by quantum mechanics $E_{\Phi_r}$, evaluated over the four measurement-setting combinations, and its standard error. The value $n=21$ is an empirical compromise between computational cost and approximation accuracy. With $s = 20$, the transition interval is $\pm\ln(9)/20 \approx \pm 0.11$; the value was chosen to ensure a smooth but selective threshold response across the tested range of $r$. All quantities are estimated from $N=100$ independent simulations, each using $10^5$ pairs of entangled photons. All reported uncertainties are standard errors of the mean (SEM), computed as $\hat{\sigma}/\sqrt{N}$ where $\hat{\sigma}$ is the sample standard deviation over the $N$ runs. The ratio $(S-2)/\mathrm{SEM}$ reported in the Results section measures how many standard errors the estimated $S$ exceeds the classical bound $S=2$; since it grows as $\sqrt{N}$, it reflects the numerical precision of the simulation rather than a property of the model.

In the second setup the source that generates photon pairs in the nonmaximally entangled state is $\ket{\Psi_r}$ (maximally entangled for $r= \pm 1$). The configuration used is $\alpha=(\SI{94.4}{\degree}, \SI{62.4}{\degree})$, $\beta=(\SI{-6.5}{\degree}, \SI{25.5}{\degree})$, $n=21$, $s=20$, with the threshold function $t(r)$ calibrated as described in Sec.~\ref{sec:calibration}. Relevant metrics include $S$ and its standard error; the single detection efficiencies $\eta_1$ and $\eta_2$ and their standard errors, averaged over the four measurement-setting combinations $(\hat{\alpha},\hat{\beta})$; the Root Mean Square Error (RMSE) between the model's correlations
$\hat{E}_{\Psi_r}$ and those predicted by quantum mechanics $E_{\Psi_r}$, evaluated over the four measurement-setting combinations, and its standard error, as in the first setup.

In the simulation programs, the Mersenne Twister algorithm is used for all random number generation (i.e., angle selection, cumulative scoring and detection mechanisms), ensuring a long period ($2^{19937}-1$) and high-quality statistical properties; the calibration procedure instead uses the deterministic PCG32 and XORWOW generators described in Sec.~\ref{sec:calibration}.

At each trial, the measurement settings $(\hat{\alpha}, \hat{\beta})$ are selected independently and uniformly at random from the two choices per station using the Mersenne Twister generator; the setting distribution is therefore not exactly balanced but subject to ordinary Monte Carlo fluctuations, which are negligible for the sample sizes used.

\section{Results}\label{sec:results}

All results reported in this section are obtained from numerical simulations; no experimental data
are used. The results apply to the specific measurement configurations and entanglement parameters described above and do not constitute a full characterization of the model over the space of all possible polarizer angles and states.

\subsection{CH--Eberhard Outcome Mapping}

Both the Clauser--Horne (CH) inequality \cite{ClauserHorne1974} and the Eberhard inequality \cite{Eberhard1993} are derived without the fair-sampling assumption. The CH--Eberhard inequality~\cite{Giustina2015} in the one-detector-per-side configuration for four measurement settings $(\hat{\alpha}_1, \hat{\alpha}_2)$ and $(\hat{\beta}_1, \hat{\beta}_2)$ is defined as
\begin{align}
J \equiv p_{++}(\hat{\alpha}_1, \hat{\beta}_1) - p_{+0}(\hat{\alpha}_1, \hat{\beta}_2) \nonumber\\
- p_{0+}(\hat{\alpha}_2, \hat{\beta}_1) - p_{++}(\hat{\alpha}_2, \hat{\beta}_2) \leq 0
\label{eq:eberhard_J}
\end{align}
where $p_{AB}(\hat{\alpha}_i, \hat{\beta}_j)=N_{ij}^{AB}/N_{ij}$ is the probability of outcomes $A$ and $B$ conditioned on the settings $(\hat{\alpha}_i, \hat{\beta}_j)$, $N_{ij}$ is the total number of trials with these settings, and $N_{ij}^{AB}$ is the number of trials with these settings and outcomes $ \{+, 0\}$ for stations 1 and 2. Under local realism the inequality $J \leq 0$ must hold.

The CH--Eberhard analysis uses a binary coarse graining of the simulation outcomes. The three-valued outcome $d_{\lambda_i} \in \{+1,-1,0\}$ of Eq.~(\ref{eq:measurement_outcome}), together with double-click events, is mapped to the two-valued CH--Eberhard outcome $\{+,0\}$ as follows: $d_{\lambda_i}=+1$ (a single click in the monitored channel) maps to ``$+$'', while $d_{\lambda_i}=-1$ (a click in the unmonitored channel), $d_{\lambda_i}=0$ (no-click), and double-click events all map to ``$0$'', i.e.\ any trial without a single click in the monitored channel.

\subsection{Numerical Simulation}

The CH--Eberhard parameter $J$ is computed from the coarse-grained outcomes. Although the simulation has access to both channels and could in principle separate double-clicks, they are assigned to ``$0$'' for consistency with this convention; their contribution is in any case negligible, since the double-click rates decrease from $\sim 10^{-3}$ at $|r|=0.1$ to $\sim 10^{-5}$--$10^{-6}$ at $|r|=1.0$  (Tables~S9--S12 in the Supplementary Material).

The model yields $J < 0$ throughout the full range of entanglement parameters and all four state configurations (Tables~\ref{tab:results_phi_r} and~\ref{tab:results_psi_r}), as required a priori for a strictly local model. The value of $J$ is therefore not a test of locality---guaranteed by construction---but the baseline against which the quantum-mechanical benchmark $J_{\rm QM}$ is compared, reported later in this section.

The CHSH values reported in the same tables are computed on the postselected detected sample under fair-sampling violation; they illustrate how a strictly local model can produce apparent CHSH-type violations when the detected sample is not representative of all generated pairs.

For the $\ket{\Phi_r^+}$ state (Table~\ref{tab:results_phi_r}) with the polarizer settings $\alpha=(\SI{0}{\degree}, \SI{45}{\degree})$ and $\beta=(\SI{22.5}{\degree}, \SI{67.5}{\degree})$, the Bell parameter $S$ increases with $r$ from $1.5690 \pm 0.0013$ at $r=0.1$ to $2.8191 \pm 0.0011$ at $r=1.0$. The ratio $(S-2)/\mathrm{SEM} \approx 740$ at $r=1.0$ reflects the numerical precision of the simulation; as discussed in the Simulation Setup, this quantity scales with $\sqrt{N}$ and should not be interpreted as a physical significance level.

For the $\ket{\Phi_r^-}$ state (Table \ref{tab:results_phi_r}) with the polarizer settings $\alpha=(\SI{0}{\degree}, \SI{45}{\degree})$ and $\beta=(\SI{22.5}{\degree}, \SI{67.5}{\degree})$, the Bell parameter $S$ decreases with increasing $|r|$, from $1.4634 \pm 0.0013$ at $r=-0.1$ to $0.0007 \pm 0.0011$ at $r=-1.0$.

\begin{table*}[hbtp]
\caption{Bell parameter $S$, joint detection efficiency $\eta$, single detection efficiencies $\eta_1$ and $\eta_2$, RMSE($\hat{E}_{\Phi_r}, E_{\Phi_r}$), CH--Eberhard parameter $J$, entanglement parameter $r$, $\alpha=(\SI{0}{\degree}, \SI{45}{\degree})$, $\beta=(\SI{22.5}{\degree}, \SI{67.5}{\degree})$.}
\label{tab:results_phi_r}
\begin{ruledtabular}
\begin{tabular}{ccccccc}
$r$ & $S$ & $\eta$ (\%) & $\eta_1$ (\%) & $\eta_2$ (\%) & RMSE & $J$ ($\times10^{-3}$) \\
\hline
$0.1$ & $1.5690 \pm 0.0013$ & $71.61 \pm 0.01$ & $81.77 \pm 0.01$ & $86.90 \pm 0.01$ & $0.0879 \pm 0.0003$ & $-599.24 \pm 0.50$ \\
$0.2$ & $1.6330 \pm 0.0013$ & $71.99 \pm 0.01$ & $83.02 \pm 0.01$ & $86.02 \pm 0.01$ & $0.1238 \pm 0.0004$ & $-562.34 \pm 0.53$ \\
$0.3$ & $1.8872 \pm 0.0012$ & $71.66 \pm 0.02$ & $83.86 \pm 0.01$ & $84.79 \pm 0.01$ & $0.0954 \pm 0.0004$ & $-513.24 \pm 0.55$ \\
$0.4$ & $2.2432 \pm 0.0011$ & $71.31 \pm 0.01$ & $84.31 \pm 0.01$ & $84.26 \pm 0.01$ & $0.0401 \pm 0.0003$ & $-483.51 \pm 0.54$ \\
$0.5$ & $2.5050 \pm 0.0012$ & $70.97 \pm 0.01$ & $84.37 \pm 0.01$ & $84.28 \pm 0.01$ & $0.0119 \pm 0.0003$ & $-473.41 \pm 0.58$ \\
$0.6$ & $2.6532 \pm 0.0011$ & $70.46 \pm 0.01$ & $84.24 \pm 0.01$ & $84.27 \pm 0.01$ & $0.0074 \pm 0.0002$ & $-470.69 \pm 0.50$ \\
$0.7$ & $2.7339 \pm 0.0010$ & $70.08 \pm 0.01$ & $84.26 \pm 0.01$ & $84.27 \pm 0.01$ & $0.0062 \pm 0.0002$ & $-475.73 \pm 0.50$ \\
$0.8$ & $2.7757 \pm 0.0011$ & $69.93 \pm 0.01$ & $84.32 \pm 0.01$ & $84.35 \pm 0.01$ & $0.0069 \pm 0.0002$ & $-481.41 \pm 0.57$ \\
$0.9$ & $2.7990 \pm 0.0011$ & $69.72 \pm 0.01$ & $84.33 \pm 0.01$ & $84.32 \pm 0.01$ & $0.0072 \pm 0.0002$ & $-487.68 \pm 0.51$ \\
$1.0$ & $2.8191 \pm 0.0011$ & $69.32 \pm 0.01$ & $84.13 \pm 0.01$ & $84.14 \pm 0.01$ & $0.0052 \pm 0.0002$ & $-495.72 \pm 0.57$ \\
$-0.1$ & $1.4634 \pm 0.0013$ & $71.62 \pm 0.01$ & $81.77 \pm 0.01$ & $86.92 \pm 0.01$ & $0.0880 \pm 0.0004$ & $-602.15 \pm 0.52$ \\
$-0.2$ & $1.2438 \pm 0.0013$ & $71.96 \pm 0.01$ & $83.01 \pm 0.01$ & $85.99 \pm 0.01$ & $0.1242 \pm 0.0004$ & $-573.80 \pm 0.52$ \\
$-0.3$ & $0.8648 \pm 0.0013$ & $71.63 \pm 0.01$ & $83.85 \pm 0.01$ & $84.78 \pm 0.01$ & $0.0960 \pm 0.0004$ & $-539.48 \pm 0.52$ \\
$-0.4$ & $0.5006 \pm 0.0012$ & $71.28 \pm 0.01$ & $84.28 \pm 0.01$ & $84.24 \pm 0.01$ & $0.0396 \pm 0.0003$ & $-507.70 \pm 0.52$ \\
$-0.5$ & $0.2724 \pm 0.0013$ & $70.95 \pm 0.01$ & $84.35 \pm 0.01$ & $84.27 \pm 0.01$ & $0.0112 \pm 0.0003$ & $-481.56 \pm 0.56$ \\
$-0.6$ & $0.1478 \pm 0.0010$ & $70.47 \pm 0.01$ & $84.26 \pm 0.01$ & $84.27 \pm 0.01$ & $0.0072 \pm 0.0002$ & $-466.41 \pm 0.51$ \\
$-0.7$ & $0.0733 \pm 0.0011$ & $70.10 \pm 0.01$ & $84.25 \pm 0.01$ & $84.29 \pm 0.01$ & $0.0064 \pm 0.0002$ & $-465.20 \pm 0.53$ \\
$-0.8$ & $0.0300 \pm 0.0011$ & $69.92 \pm 0.01$ & $84.33 \pm 0.01$ & $84.34 \pm 0.01$ & $0.0074 \pm 0.0003$ & $-470.96 \pm 0.51$ \\
$-0.9$ & $0.0073 \pm 0.0010$ & $69.72 \pm 0.01$ & $84.33 \pm 0.01$ & $84.33 \pm 0.01$ & $0.0071 \pm 0.0002$ & $-481.41 \pm 0.54$ \\
$-1.0$ & $0.0007 \pm 0.0011$ & $69.30 \pm 0.01$ & $84.13 \pm 0.01$ & $84.13 \pm 0.01$ & $0.0054 \pm 0.0002$ & $-495.26 \pm 0.48$ \\
\end{tabular}
\end{ruledtabular}
\end{table*}

For the $\ket{\Psi_r^+}$ state (Table \ref{tab:results_psi_r}) with the polarizer settings $\alpha=(\SI{94.4}{\degree}, \SI{62.4}{\degree})$ and $\beta=(\SI{-6.5}{\degree}, \SI{25.5}{\degree})$, the Bell parameter $S$ ranges from $1.4628 \pm 0.0010$ at $r=0.1$ to $1.8308 \pm 0.0010$ at $r=1.0$.

For the $\ket{\Psi_r^-}$ state (Table \ref{tab:results_psi_r}) with the polarizer settings $\alpha=(\SI{94.4}{\degree}, \SI{62.4}{\degree})$ and $\beta=(\SI{-6.5}{\degree}, \SI{25.5}{\degree})$, the Bell parameter $S$ ranges from $1.2925 \pm 0.0011$ at $r=-0.1$ to $0.6663 \pm 0.0010$ at $r=-1.0$.

For the $\ket{\Phi_r^+}$ state at $r=1.0$, the single-detector efficiencies are $\eta_1 = 84.13\% \pm 0.01\%$ and $\eta_2 = 84.14\% \pm 0.01\%$, and the joint detection efficiency is $\eta = 69.32\% \pm 0.01\%$.

The model achieves the highest accuracy at $|r|=1$, with respect to both the predicted quantum correlations and the coincidence probabilities.

Mermin's bound of $82.8\%$~\cite{Mermin1986} is a threshold on
single-detector efficiency derived under two conditions. The first is
that the probability of registering an event at one detector when the
other detector fails to register a count, and the probability of
registering an event at a detector without regard to what is happening
at the other detector, are both independent of the settings of the
polarizers. The second is that the detection probabilities are entirely
determined by the quantum efficiency of the detectors, so that the ratio
of the number of detected pairs to the number of detected pairs and
single events is $\eta_s/(2-\eta_s)$, where $\eta_s$ is the single-detector efficiency; the same threshold follows from the analysis of Garg and Mermin~\cite{GargMermin1987}. The present
model violates the first condition for $|r|<1$, where the single-detector
detection efficiencies are setting-dependent, and the second at every $r$, the two
detections being correlated through the hidden variable: at $|r|=1$ the
ratio is $\eta/(\eta_1+\eta_2-\eta)=0.70$ rather than $\eta_s/(2-\eta_s)=0.73$
(Table~\ref{tab:results_phi_r}). The bound therefore does not apply.

\begin{figure}[htbp]
\centering
\includegraphics[width=\columnwidth]{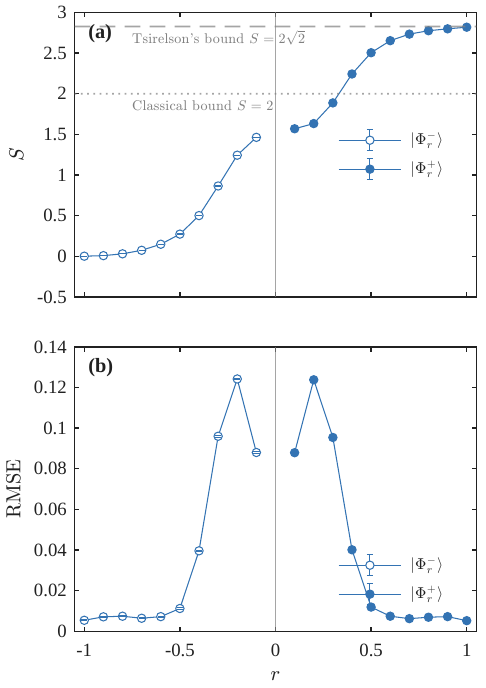}
\caption{(a) Bell parameter $S$ and (b) RMSE between model and quantum-mechanical correlations as functions of $r$, for states $\ket{\Phi_r^+}$ and $\ket{\Phi_r^-}$, $\alpha=(\SI{0}{\degree}, \SI{45}{\degree})$, $\beta=(\SI{22.5}{\degree}, \SI{67.5}{\degree})$. Error bars (standard error of the mean
over 100 independent runs) are smaller than the marker size.}
\label{fig:S_RMSE_Phi}
\end{figure}

\begin{figure}[htbp]
\centering
\includegraphics[width=\columnwidth]{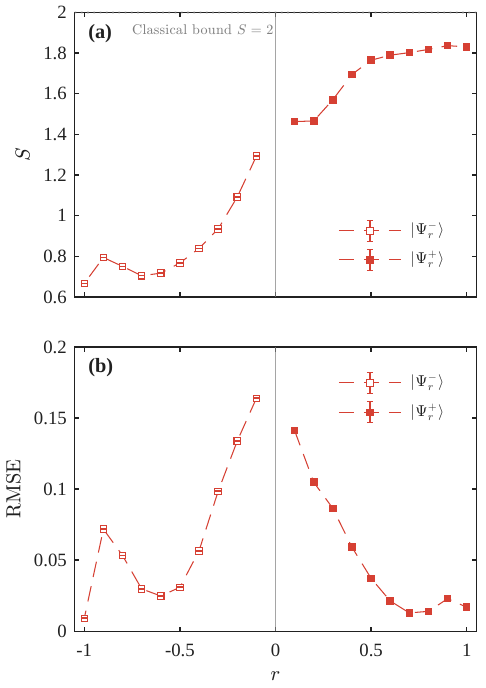}
\caption{(a) Bell parameter $S$ and (b) RMSE between model and quantum-mechanical correlations as functions of $r$, for states $\ket{\Psi_r^+}$ and $\ket{\Psi_r^-}$, $\alpha=(\SI{94.4}{\degree}, \SI{62.4}{\degree})$, $\beta=(\SI{-6.5}{\degree}, \SI{25.5}{\degree})$. Error bars (standard error of the mean over 100 independent runs) are smaller than the marker size.}
\label{fig:S_RMSE_Psi}
\end{figure}

Figures~\ref{fig:S_RMSE_Phi} and~\ref{fig:S_RMSE_Psi} show the Bell parameter $S$ and the RMSE between the model's correlations and the quantum mechanical correlations, as functions of $r$ for the states $\ket{\Phi_r}$ and $\ket{\Psi_r}$ respectively, providing a visual summary of the data reported in Tables~\ref{tab:results_phi_r}--\ref{tab:results_psi_r}.

\begin{figure*}[htbp]
\centering
\includegraphics[width=\textwidth]{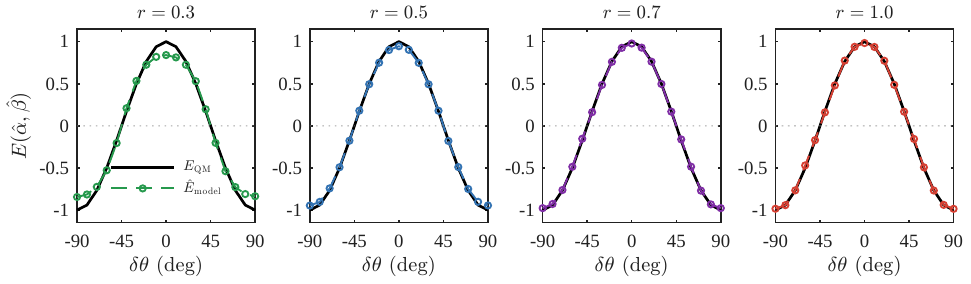}
\caption{Model correlations $\hat{E}$ (markers) versus quantum-mechanical correlations $E_\mathrm{QM}$ (solid) as a function of the relative analyzer angle $\delta\theta$, for $r \in \{0.3, 0.5, 0.7, 1.0\}$, for the state $\ket{\Phi_r^+}$, with $\hat{\alpha} =
0^\circ$ fixed.}
\label{fig:correlations_phi}
\end{figure*}

\begin{figure*}[htbp]
\centering
\includegraphics[width=\textwidth]{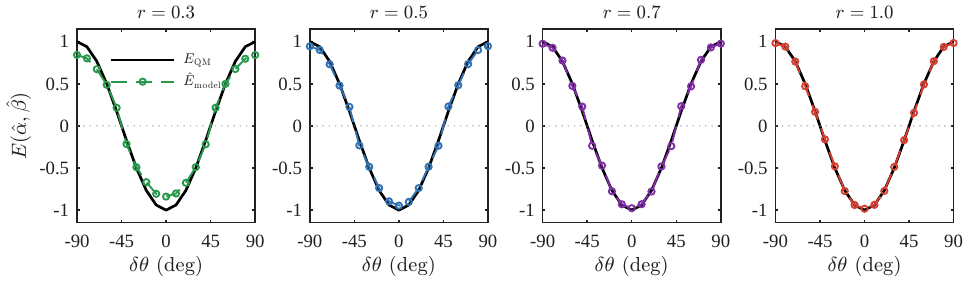}
\caption{Model correlations $\hat{E}$ (markers) versus quantum-mechanical correlations $E_\mathrm{QM}$ (solid) as a function of the relative analyzer angle $\delta\theta$, for $r \in \{0.3, 0.5, 0.7, 1.0\}$, for the state $\ket{\Psi_r^+}$, with $\hat{\alpha} =
0^\circ$ fixed.}
\label{fig:correlations_psi}
\end{figure*}

Figures~\ref{fig:correlations_phi} and~\ref{fig:correlations_psi} show the model correlations against the quantum-mechanical correlations as a function of the relative analyzer angle $\delta\theta$, for the states $\ket{\Phi_r}$ and $\ket{\Psi_r}$ respectively, for $r \in \{0.3, 0.5, 0.7, 1.0\}$, with $\hat{\alpha} = 0^\circ$ fixed. The agreement improves systematically as $|r| \to 1$.

\begin{figure}[htbp]
\centering
\includegraphics[width=\columnwidth]{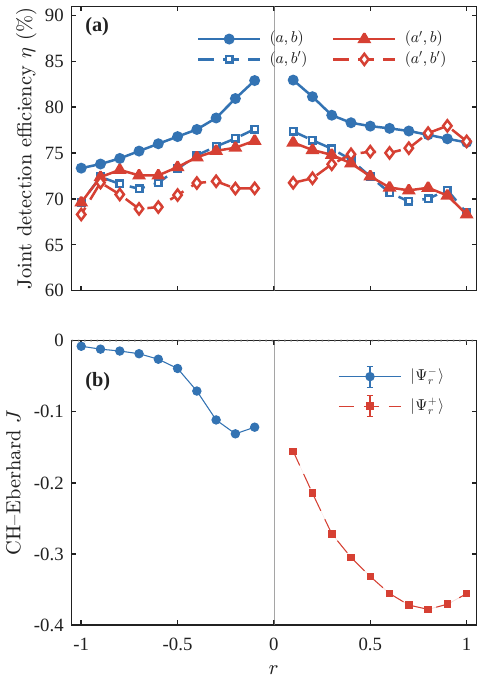}
\caption{(a) Setting-dependent joint detection efficiency for the fixed-threshold model $\ket{\Psi_r}$, $\alpha=(\SI{94.4}{\degree}, \SI{62.4}{\degree})$, $\beta=(\SI{-6.5}{\degree}, \SI{25.5}{\degree})$, Table~S4 in the Supplementary Material. (b) CH--Eberhard parameter $J$ for the fixed-threshold model $\ket{\Psi_r}$, Table~\ref{tab:results_psi_r}.}
\label{fig:diagnostics_psi}
\end{figure}

Tables~S3 and~S4 in the Supplementary Material and Figs.~\ref{fig:diagnostics_phi}a, \ref{fig:diagnostics_psi}a report the joint detection efficiency, computed for each of the four measurement-setting combinations $(\hat{\alpha},\hat{\beta})$, averaged over $N=100$ independent runs; the results show that it varies across settings throughout the full range of entanglement parameters. For the $|\Phi_r\rangle$ states, the curves for $(a,b)$ and $(a,b')$ overlap, as do those for $(a',b)$ and $(a',b')$.

The no-signaling condition requires that the marginal detection probability at each station be independent of the remote station's setting. Defining $|\Delta P_A^{(a)}| = |P(A|a,b)-P(A|a,b')|$ as the variation of the marginal detection probability at station~A under a change of the remote setting of station~B, and analogously for $|\Delta P_A^{(a')}|$, $|\Delta P_B^{(b)}|$, and $|\Delta P_B^{(b')}|$, these deviations are reported with their standard errors and $Z$-scores [$Z=|\Delta P|/\mathrm{SE}(|\Delta P|)$] in the Supplementary Material (Tables~S5--S6). They all remain below $9\times10^{-4}$ and are individually consistent with zero at the finite-sample level: $Z<2$ for all but four ($\ket{\Phi_r}$) and five ($\ket{\Psi_r}$) of the eighty entries per state, no deviation exceeds $2.6$ standard errors, and no systematic dependence on the remote setting is present, consistent with locality.

The analogous diagnostic on the postselected outcome marginals is reported with standard errors and $Z$-scores in the Supplementary Material (Tables~S7--S8): $P_{\rm ps}(A{=}{+1}\mid a,b)$ and $P_{\rm ps}(B{=}{+1}\mid a,b)$, the fractions of detected events with outcome $+1$, must likewise be independent of the remote setting, so that $|\Delta P_{\rm ps}(A\mid a)| = |P_{\rm ps}(A{=}{+1}\mid a,b) - P_{\rm ps}(A{=}{+1}\mid a,b')|$ and the analogous station-B quantities vanish up to finite-sample fluctuations; all deviations remain below $1.8\times10^{-3}$ for $\ket{\Phi_r}$ and below $1.1\times10^{-3}$ for $\ket{\Psi_r}$, with a single entry reaching $Z\approx3.4$, confirming no-signaling also at the outcome level within the detected sample.

Across both states and both diagnostics, $19$ of the $320$ deviations exceed $Z=2$, in line with the ${\approx}15$ expected by chance under the no-signaling null, with no systematic dependence on $r$, station, or setting.

For each measurement setting, Tables~S9--S12 in the Supplementary Material report the double-click rate (both channels fire simultaneously) and the no-click rate (neither channel fires). Non-detection is dominated by no-click events: double-click rates are of order $10^{-3}$ at $|r|=0.1$ and decrease to $\sim10^{-5}$--$10^{-6}$ at $|r|=1.0$, while no-click rates remain in the range $8$--$27\%$.

\begin{figure}[htbp]
\centering
\includegraphics[width=\columnwidth]{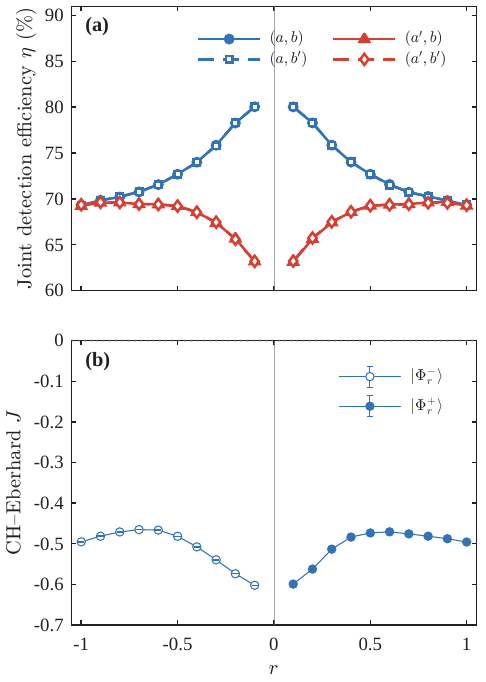}
\caption{(a) Setting-dependent joint detection efficiency for the fixed-threshold model $\ket{\Phi_r}$, $\alpha=(\SI{0}{\degree}, \SI{45}{\degree})$, $\beta=(\SI{22.5}{\degree}, \SI{67.5}{\degree})$, Table~S3 in the Supplementary Material. The curves for $(a,b)$ and $(a,b')$ overlap, as do those for $(a',b)$ and $(a',b')$. (b) CH--Eberhard parameter $J$ for the fixed-threshold model $\ket{\Phi_r}$, Table~\ref{tab:results_phi_r}.}
\label{fig:diagnostics_phi}
\end{figure}

Tables~S1 and~S2 in the Supplementary Material report the RMSE between the model's correlations and the quantum mechanical predictions, averaged over an $18\times18$ uniform grid of polarizer angles $(\hat{\alpha},\hat{\beta})$ with step $\SI{10}{\degree}$ in $[\SI{0}{\degree},\SI{170}{\degree}]$, from 100 independent runs of $10^4$ pairs each. For the $\ket{\Phi_r}$ state, RMSE ranges from $\approx 0.20$ at $|r|=0.1$ to $\approx 0.014$ at $|r|=1.0$, and for the $\ket{\Psi_r}$ state from $\approx 0.15$ at $|r|=0.1$ to $\approx 0.014$ at $|r|=1.0$.

\begin{table*}[hbtp]
\caption{Bell parameter $S$, joint detection efficiency $\eta$, single detection efficiencies $\eta_1$ and $\eta_2$, RMSE($\hat{E}_{\Psi_r}, E_{\Psi_r}$), CH--Eberhard parameter $J$, entanglement parameter $r$, $\alpha=(\SI{94.4}{\degree}, \SI{62.4}{\degree})$, $\beta=(\SI{-6.5}{\degree}, \SI{25.5}{\degree})$.}
\label{tab:results_psi_r}
\begin{ruledtabular}
\begin{tabular}{ccccccc}
$r$ & $S$ & $\eta$ (\%) & $\eta_1$ (\%) & $\eta_2$ (\%) & RMSE & $J$ ($\times10^{-3}$) \\
\hline
$0.1$ & $1.4628 \pm 0.0010$ & $77.05 \pm 0.01$ & $87.07 \pm 0.01$ & $87.45 \pm 0.01$ & $0.1411 \pm 0.0002$ & $-155.7 \pm 0.3$ \\
$0.2$ & $1.4663 \pm 0.0010$ & $76.27 \pm 0.01$ & $86.74 \pm 0.01$ & $86.97 \pm 0.01$ & $0.1048 \pm 0.0002$ & $-214.7 \pm 0.3$ \\
$0.3$ & $1.5705 \pm 0.0010$ & $75.79 \pm 0.01$ & $86.52 \pm 0.01$ & $86.64 \pm 0.01$ & $0.0864 \pm 0.0002$ & $-271.7 \pm 0.5$ \\
$0.4$ & $1.6952 \pm 0.0010$ & $75.32 \pm 0.01$ & $86.24 \pm 0.01$ & $86.29 \pm 0.01$ & $0.0591 \pm 0.0002$ & $-304.9 \pm 0.5$ \\
$0.5$ & $1.7652 \pm 0.0011$ & $74.49 \pm 0.01$ & $85.72 \pm 0.01$ & $85.60 \pm 0.01$ & $0.0372 \pm 0.0002$ & $-331.8 \pm 0.6$ \\
$0.6$ & $1.7899 \pm 0.0010$ & $73.66 \pm 0.01$ & $85.24 \pm 0.01$ & $84.83 \pm 0.01$ & $0.0213 \pm 0.0002$ & $-355.7 \pm 0.5$ \\
$0.7$ & $1.8025 \pm 0.0011$ & $73.39 \pm 0.01$ & $85.14 \pm 0.01$ & $84.43 \pm 0.01$ & $0.0129 \pm 0.0002$ & $-372.0 \pm 0.5$ \\
$0.8$ & $1.8182 \pm 0.0010$ & $73.85 \pm 0.01$ & $85.39 \pm 0.01$ & $84.65 \pm 0.01$ & $0.0140 \pm 0.0002$ & $-378.0 \pm 0.5$ \\
$0.9$ & $1.8367 \pm 0.0011$ & $73.96 \pm 0.01$ & $85.09 \pm 0.01$ & $85.27 \pm 0.01$ & $0.0229 \pm 0.0003$ & $-370.8 \pm 0.5$ \\
$1.0$ & $1.8308 \pm 0.0010$ & $72.30 \pm 0.01$ & $84.15 \pm 0.01$ & $84.11 \pm 0.01$ & $0.0169 \pm 0.0003$ & $-356.0 \pm 0.5$ \\
$-0.1$ & $1.2925 \pm 0.0011$ & $77.00 \pm 0.01$ & $87.10 \pm 0.01$ & $87.47 \pm 0.01$ & $0.1638 \pm 0.0003$ & $-121.8 \pm 0.3$ \\
$-0.2$ & $1.0917 \pm 0.0012$ & $76.06 \pm 0.01$ & $86.73 \pm 0.01$ & $86.95 \pm 0.01$ & $0.1338 \pm 0.0003$ & $-131.1 \pm 0.3$ \\
$-0.3$ & $0.9341 \pm 0.0011$ & $75.43 \pm 0.01$ & $86.51 \pm 0.01$ & $86.64 \pm 0.01$ & $0.0984 \pm 0.0003$ & $-111.6 \pm 0.3$ \\
$-0.4$ & $0.8385 \pm 0.0011$ & $74.65 \pm 0.01$ & $86.23 \pm 0.01$ & $86.26 \pm 0.01$ & $0.0565 \pm 0.0003$ & $-70.9 \pm 0.4$ \\
$-0.5$ & $0.7669 \pm 0.0011$ & $73.49 \pm 0.01$ & $85.75 \pm 0.01$ & $85.57 \pm 0.01$ & $0.0309 \pm 0.0003$ & $-39.2 \pm 0.5$ \\
$-0.6$ & $0.7166 \pm 0.0010$ & $72.36 \pm 0.01$ & $85.25 \pm 0.01$ & $84.83 \pm 0.01$ & $0.0248 \pm 0.0003$ & $-26.2 \pm 0.4$ \\
$-0.7$ & $0.7041 \pm 0.0011$ & $71.94 \pm 0.01$ & $85.15 \pm 0.01$ & $84.43 \pm 0.01$ & $0.0298 \pm 0.0003$ & $-18.4 \pm 0.5$ \\
$-0.8$ & $0.7501 \pm 0.0010$ & $72.42 \pm 0.01$ & $85.36 \pm 0.01$ & $84.65 \pm 0.01$ & $0.0533 \pm 0.0003$ & $-14.6 \pm 0.5$ \\
$-0.9$ & $0.7937 \pm 0.0012$ & $72.59 \pm 0.01$ & $85.09 \pm 0.01$ & $85.27 \pm 0.01$ & $0.0718 \pm 0.0004$ & $-12.0 \pm 0.6$ \\
$-1.0$ & $0.6663 \pm 0.0010$ & $70.22 \pm 0.01$ & $84.13 \pm 0.01$ & $84.14 \pm 0.01$ & $0.0091 \pm 0.0003$ & $-7.9 \pm 0.5$ \\
\end{tabular}
\end{ruledtabular}
\end{table*}

The quantum joint detection probabilities for $\ket{\Psi_r}$ entering the CH--Eberhard analysis are obtained from the Born rule. Defining the projection
$\ket{\widetilde\Psi_{++}} \equiv
\bigl(\hat{\Pi}_+(\hat{\alpha})\otimes\hat{\Pi}_+(\hat{\beta})\bigr)\ket{\Psi_r}$,
applying the projectors term by term to
$\ket{\Psi_r}=(\ket{HV}+r\ket{VH})/\sqrt{1+r^2}$ gives
\begin{equation}
\ket{\widetilde\Psi_{++}} =
\frac{\cos\hat{\alpha}\sin\hat{\beta}+r\sin\hat{\alpha}\cos\hat{\beta}}{\sqrt{1+r^2}}\,
\ket{\hat\alpha}\ket{\hat{\beta}},
\end{equation}
and $\|\ket{\widetilde\Psi_{++}}\|^2$ gives $P_{++}^{\ket{\Psi}}$ of Eq.~\ref{eq:psi_prob_pp}.
The remaining three probabilities follow the same way, with the
$\hat\Pi_-$ relations replacing $\hat\Pi_+$ at whichever station carries
a $-$ outcome label (both stations for $P_{--}$, one station for
$P_{+-}$ and $P_{-+}$). The four coincidence probabilities for $\ket{\Psi_r}$ are
therefore
\begin{align}
P_{++}^{\ket{\Psi}}(\hat{\alpha}, \hat{\beta}, r) &= \frac{\left(\cos(\hat{\alpha})\sin(\hat{\beta}) + r\sin(\hat{\alpha})\cos(\hat{\beta})\right)^2}{1+r^2} \label{eq:psi_prob_pp} \\
P_{--}^{\ket{\Psi}}(\hat{\alpha}, \hat{\beta}, r) &= \frac{\left(\sin(\hat{\alpha})\cos(\hat{\beta}) + r\cos(\hat{\alpha})\sin(\hat{\beta})\right)^2}{1+r^2} \label{eq:psi_prob_mm} \\
P_{+-}^{\ket{\Psi}}(\hat{\alpha}, \hat{\beta}, r) &= \frac{\left(\cos(\hat{\alpha})\cos(\hat{\beta}) - r\sin(\hat{\alpha})\sin(\hat{\beta})\right)^2}{1+r^2} \label{eq:psi_prob_pm} \\
P_{-+}^{\ket{\Psi}}(\hat{\alpha}, \hat{\beta}, r) &= \frac{\left(\sin(\hat{\alpha})\sin(\hat{\beta}) - r\cos(\hat{\alpha})\cos(\hat{\beta})\right)^2}{1+r^2} \label{eq:psi_prob_mp}
\end{align}

The quantum-mechanical prediction $J_{\rm QM}$ is then evaluated from the CH--Eberhard parameter of Eq.~(\ref{eq:eberhard_J}), in the same one-detector-per-side $\{+,0\}$ outcome convention and normalized to the model's single-detection efficiencies $\eta_1$ and $\eta_2$, so that both quantities are constructed under identical conditions. A $+$ outcome at a station requires the photon to be detected---with probability $\eta_1$ at station~1 or $\eta_2$ at station~2---and to produce a click in the monitored channel, while $0$ collects every other case. The outcome probabilities entering Eq.~(\ref{eq:eberhard_J}) are therefore
\begin{align}
p_{++}(\hat{\alpha},\hat{\beta}) &= \eta_1\eta_2\, P_{++}^{\ket{\Psi}}(\hat{\alpha},\hat{\beta}), \\
p_{+0}(\hat{\alpha},\hat{\beta}) &= \eta_1\bigl(P_{++}^{\ket{\Psi}}(\hat{\alpha},\hat{\beta}) + P_{+-}^{\ket{\Psi}}(\hat{\alpha},\hat{\beta})\bigr) - \eta_1\eta_2\, P_{++}^{\ket{\Psi}}(\hat{\alpha},\hat{\beta}), \\
p_{0+}(\hat{\alpha},\hat{\beta}) &= \eta_2\bigl(P_{++}^{\ket{\Psi}}(\hat{\alpha},\hat{\beta}) + P_{-+}^{\ket{\Psi}}(\hat{\alpha},\hat{\beta})\bigr) - \eta_1\eta_2\, P_{++}^{\ket{\Psi}}(\hat{\alpha},\hat{\beta}),
\end{align}
where the sums $P_{++}^{\ket{\Psi}} + P_{+-}^{\ket{\Psi}}$ and $P_{++}^{\ket{\Psi}} + P_{-+}^{\ket{\Psi}}$ are the single-station probabilities of a $+$ outcome at stations~1 and~2, obtained from the coincidence probabilities (\ref{eq:psi_prob_pp})--(\ref{eq:psi_prob_mp}) by summing over the unobserved outcome at the remote station. Substituting Eqs.~(\ref{eq:psi_prob_pp})--(\ref{eq:psi_prob_mp}), these reduce to
\begin{align}
P_{++}^{\ket{\Psi}} + P_{+-}^{\ket{\Psi}} &= \frac{\cos^2\hat{\alpha} + r^2\sin^2\hat{\alpha}}{1+r^2}, \\
P_{++}^{\ket{\Psi}} + P_{-+}^{\ket{\Psi}} &= \frac{\sin^2\hat{\beta} + r^2\cos^2\hat{\beta}}{1+r^2},
\end{align}
which are independent of the remote setting, as required by no-signaling. Inserting $p_{++}$, $p_{+0}$, and $p_{0+}$ into Eq.~(\ref{eq:eberhard_J}) at the four setting pairs $(\hat{\alpha}_i,\hat{\beta}_j)$ yields $J_{\rm QM}$ normalized to the model's single-detection efficiencies $\eta_1$ and $\eta_2$. In this construction $J_{\rm QM}$ treats the model's setting-averaged efficiencies as setting-independent constants; the setting dependence of the detection efficiency, which is the operative loophole mechanism, enters $J_{\rm model}$ only.

For the $\ket{\Phi_r}$ state at the CHSH-optimal angles $\alpha=(\SI{0}{\degree}, \SI{45}{\degree})$, $\beta=(\SI{22.5}{\degree}, \SI{67.5}{\degree})$, $J_{\rm QM}<0$ throughout the full range $r \in [-1,1]$: both quantum mechanics and the model predict $J<0$, so the comparison is inconclusive for this state and the tested angle configuration does not provide a nontrivial CH--Eberhard test.

For the $\ket{\Psi_r}$ state at the tested angles $\alpha=(\SI{94.4}{\degree}, \SI{62.4}{\degree})$, $\beta=(\SI{-6.5}{\degree}, \SI{25.5}{\degree})$ the situation differs. For $r>0$ both $J_{\rm model}$ and $J_{\rm QM}$ are negative, again a trivial comparison; for $r<0$, however, $J_{\rm QM}>0$ for $r\in[-0.9,-0.2]$, reaching a maximum of $J_{\rm QM}\approx +4.3\times10^{-2}$ at $r=-0.5$, while $J_{\rm model}<0$ throughout. This is the regime of nontrivial comparison (Table~\ref{tab:J_comparison_psi_r}): at the tested efficiencies and angles, quantum mechanics violates the CH--Eberhard inequality ($J_{\rm QM}>0$) while the local model does not ($J_{\rm model}<0$), and the gap $J_{\rm QM}-J_{\rm model}$ measures how far the local model remains from the quantum prediction precisely where the CH--Eberhard test is discriminating. At $r=-0.1$ and $r=-1.0$, $J_{\rm QM}$ is again negative, so among the sampled values the nontrivial regime is confined to $r\in[-0.9,-0.2]$.

\begin{table}[htbp]
\caption{CH--Eberhard parameter of the model ($J_{\rm model}$) and the quantum-mechanical prediction ($J_{\rm QM}$) evaluated at the same mean station 1 and 2 detection efficiencies as the model, in the one-detector-per-side $\{+,0\}$ outcome convention, for $\ket{\Psi_r}$, $\alpha=(\SI{94.4}{\degree}, \SI{62.4}{\degree})$, $\beta=(\SI{-6.5}{\degree}, \SI{25.5}{\degree})$.}
\label{tab:J_comparison_psi_r}
\begin{ruledtabular}
\begin{tabular}{ccc}
$r$ & $J_{\rm model} (\times 10^{-3})$ & $J_{\rm QM} (\times 10^{-3})$ \\
\hline
$0.1$ & $-155.66 \pm 0.32$ & $-80.00$ \\
$0.2$ & $-214.66 \pm 0.31$ & $-117.74$ \\
$0.3$ & $-271.71 \pm 0.52$ & $-155.51$ \\
$0.4$ & $-304.92 \pm 0.49$ & $-191.32$ \\
$0.5$ & $-331.81 \pm 0.59$ & $-223.71$ \\
$0.6$ & $-355.74 \pm 0.53$ & $-252.67$ \\
$0.7$ & $-371.97 \pm 0.50$ & $-278.11$ \\
$0.8$ & $-378.02 \pm 0.53$ & $-299.34$ \\
$0.9$ & $-370.79 \pm 0.53$ & $-316.50$ \\
$1.0$ & $-355.97 \pm 0.52$ & $-332.69$ \\
$-0.1$ & $-121.75 \pm 0.26$ & $-11.59$ \\
$-0.2$ & $-131.09 \pm 0.34$ & $+13.87$ \\
$-0.3$ & $-111.55 \pm 0.32$ & $+31.69$ \\
$-0.4$ & $-70.94 \pm 0.44$ & $+41.46$ \\
$-0.5$ & $-39.18 \pm 0.46$ & $+42.62$ \\
$-0.6$ & $-26.15 \pm 0.42$ & $+36.84$ \\
$-0.7$ & $-18.42 \pm 0.45$ & $+28.33$ \\
$-0.8$ & $-14.64 \pm 0.52$ & $+20.49$ \\
$-0.9$ & $-11.97 \pm 0.55$ & $+10.85$ \\
$-1.0$ & $-7.90 \pm 0.46$ & $-11.55$ \\
\end{tabular}
\end{ruledtabular}
\end{table}

\section{Adaptive-Threshold Model}
\label{sec:adaptive_threshold}

\begin{table*}[hbtp]
\centering
\setlength{\tabcolsep}{3.5pt}
\caption{Adaptive-threshold model: Bell parameter $S$, joint detection efficiency $\eta$, single detection efficiencies $\eta_1$ and $\eta_2$, RMSE($\hat{E}_{\Psi_r^-}, E_{\Psi_r^-}$), CH--Eberhard parameter $J$, $\alpha=(\SI{94.4}{\degree}, \SI{62.4}{\degree})$, $\beta=(\SI{-6.5}{\degree}, \SI{25.5}{\degree})$, $-1.0 \leq r \leq -0.1$.}
\label{tab:memory_results}
\begin{ruledtabular}
\begin{tabular}{ccccccc}
$r$ & $S$ & $\eta$ (\%) & $\eta_1$ (\%) & $\eta_2$ (\%) & RMSE &
$J$ ($\times 10^{-3}$) \\
\hline
$-0.1$ & $1.2152 \pm 0.0009$ & $83.83 \pm 0.01$ & $91.14 \pm 0.01$ & $91.58 \pm 0.01$ & $0.1549 \pm 0.0003$ & $-130.47 \pm 0.31$ \\
$-0.2$ & $1.0207 \pm 0.0010$ & $83.27 \pm 0.01$ & $90.96 \pm 0.01$ & $91.23 \pm 0.01$ & $0.1454 \pm 0.0003$ & $-134.48 \pm 0.32$ \\
$-0.3$ & $0.8711 \pm 0.0010$ & $83.17 \pm 0.01$ & $91.02 \pm 0.01$ & $91.17 \pm 0.01$ & $0.1249 \pm 0.0003$ & $-109.69 \pm 0.39$ \\
$-0.4$ & $0.7869 \pm 0.0012$ & $82.72 \pm 0.01$ & $90.87 \pm 0.01$ & $90.92 \pm 0.01$ & $0.0899 \pm 0.0002$ & $-63.85 \pm 0.46$ \\
$-0.5$ & $0.7224 \pm 0.0011$ & $81.34 \pm 0.01$ & $90.22 \pm 0.01$ & $90.14 \pm 0.01$ & $0.0723 \pm 0.0003$ & $-35.06 \pm 0.49$ \\
$-0.6$ & $0.6783 \pm 0.0010$ & $79.90 \pm 0.01$ & $89.56 \pm 0.01$ & $89.24 \pm 0.01$ & $0.0693 \pm 0.0002$ & $-23.77 \pm 0.47$ \\
$-0.7$ & $0.6657 \pm 0.0011$ & $79.14 \pm 0.01$ & $89.27 \pm 0.01$ & $88.67 \pm 0.01$ & $0.0695 \pm 0.0003$ & $-16.60 \pm 0.40$ \\
$-0.8$ & $0.7112 \pm 0.0010$ & $79.30 \pm 0.01$ & $89.34 \pm 0.01$ & $88.69 \pm 0.01$ & $0.0753 \pm 0.0003$ & $-14.62 \pm 0.45$ \\
$-0.9$ & $0.7736 \pm 0.0011$ & $79.38 \pm 0.01$ & $89.01 \pm 0.01$ & $89.15 \pm 0.01$ & $0.0878 \pm 0.0003$ & $-11.85 \pm 0.45$ \\
$-1.0$ & $0.6540 \pm 0.0010$ & $77.20 \pm 0.01$ & $88.09 \pm 0.01$ & $88.08 \pm 0.01$ & $0.0508 \pm 0.0002$ & $-6.86 \pm 0.45$ \\
\end{tabular}
\end{ruledtabular}
\end{table*}

This section investigates whether a specific adaptive-threshold rule, implementing the one-sided memory mechanism---as defined by
Barrett et al.~\cite{Barrett2002memory}, where each detector maintains an independent local history with no inter-station communication---produces a systematic shift of the CH--Eberhard parameter $J$ relative to the fixed-threshold model. The adaptive mechanism adjusts each detector's discrimination threshold based on accumulated local measurement history, and the CH--Eberhard parameter $J$ is used to assess whether this produces any systematic effect under the tested conditions.

The adaptive mechanism follows a recursive estimation scheme: each station maintains running estimates---implemented as exponential moving averages---of the mean magnitude and variance of its local
net score for each measurement setting, uses these to compute a signal-to-noise ratio, and adjusts its detection threshold accordingly.

Each measurement station maintains an independent memory state for each of its two settings. When a setting is selected, the corresponding memory state is updated using only local information (the net score for that measurement), with no communication between stations. For a given setting at station $i$, running estimates of the mean magnitude $\mu_i^{(k)}$ and variance $v_i^{(k)}$ of the channel imbalance are maintained
\begin{align}
\mu_i^{(k)} &= (1-\alpha_{\mu}^{(k)})\,\mu_i^{(k-1)} + \alpha_{\mu}^{(k)}\,|p_i^{(k)}|
\label{eq:running_mean}\\
v_i^{(k)} &= (1-\alpha_v)\,v_i^{(k-1)} + \alpha_v\,\bigl(|p_i^{(k)}| - \mu_i^{(k)}\bigr)^2
\label{eq:running_variance}
\end{align}
where $k$ denotes the number of measurement events recorded for that setting, $\mu_i^{(k)}$ in Eq.~(\ref{eq:running_variance}) denotes the updated mean from Eq.~(\ref{eq:running_mean}), $\alpha_{\mu}^{(k)}$ is an adaptive learning rate, $\alpha_v = 0.04$ is a fixed learning rate for variance, and the initialization is $\mu_i^{(1)} = |p_i^{(1)}|$ and $v_i^{(1)} = 0$. The net scores $p_i^{(k)}$ are derived from Eqs.~(\ref{eq:net_scores_1})-(\ref{eq:net_scores_2}) as

\begin{align}
p_1^{(k)} = (\hat{c}_1^+ - \hat{c}_1^-), \qquad
p_2^{(k)} = \sigma_{\mathcal{B}}(\hat{c}_2^+ - \hat{c}_2^-)
\end{align}

Variance-aware memory is implemented by modulating the learning rate
$\alpha_{\mu}^{(k)}$ inversely with the local variance
\begin{align}
k_i^{(k)} &=
\left[
\frac{V_{\mathrm{ref}}}{v_i^{(k)}+\varepsilon}
\right]_{k_{\min}}^{k_{\max}} \\
\alpha_{\mu}^{(k)} &=
\left[
\alpha_{\mu,0}\,k_i^{(k)}
\right]_{\alpha_{\mu,\min}}^{\alpha_{\mu,\max}}
\end{align}
where $[x]_{a}^{b} = \max(a,\min(x,b))$ denotes projection onto the interval $[a,b]$, $V_{\text{ref}} = 0.015$ sets the reference variance scale, $\alpha_{\mu,0} = 0.01$ is the base learning rate, $k_{\min} = 0.3$ and $k_{\max} = 3.0$ bound the modulation factor, $\alpha_{\mu,\min} = 0.003$ and $\alpha_{\mu,\max} = 0.06$ bound the effective learning rate, and $\varepsilon = 10^{-6}$ is a regularization constant. Large local variance leads to smaller $\alpha_{\mu}^{(k)}$ (longer memory), allowing the mean estimate to stabilize; small variance permits faster adaptation.

The memory state for each setting is initialized with $\mu_i^{(1)} = |p_i^{(1)}|$ and $v_i^{(1)} = 0$
at the first measurement event for that setting. During the warm-up period of $K_{\text{w}} = 1500$
measurements per setting, the threshold remains fixed at the base value $t(r)$ while the running
estimates $\mu_i^{(k)}$ and $v_i^{(k)}$ accumulate; adaptation begins only after warm-up is complete. The signal-to-noise ratio is defined as
\begin{equation}
\mathrm{SNR}_i^{(k)} = \frac{\mu_i^{(k)}}{\sqrt{v_i^{(k)} + \varepsilon'}}
\label{eq:snr}
\end{equation}
where $\varepsilon' = 10^{-9}$ is a regularization constant. The adaptation strength $A_i^{(k)}$ is modulated based on SNR
\begin{equation}
A_i^{(k)} =
\begin{cases}
A_0,
& \mathrm{SNR}_i^{(k)} > \mathrm{SNR}_{\text{high}}\\
0.7\,A_0,
& \mathrm{SNR}_{\text{low}} < \mathrm{SNR}_i^{(k)}
\le \mathrm{SNR}_{\text{high}}\\
0.4\,A_0,
& \mathrm{SNR}_i^{(k)} \le \mathrm{SNR}_{\text{low}}
\end{cases}
\label{eq:adaptation_strength}
\end{equation}
with $A_0 = 0.40$ and thresholds $\mathrm{SNR}_{\text{low}} = 7.0$, $\mathrm{SNR}_{\text{high}} = 12.0$. The adaptive threshold $t_i^{(k)}$ at station $i$ for a given setting is computed by first obtaining an unconstrained update
\begin{equation}
\tau_i^{(k)} =
t(r)\!\left[1 - A_i^{(k)}\bigl(\chi_i^{(k)} - 1\bigr)\right]
\label{eq:unconstrained_threshold}
\end{equation}
where $\chi_i^{(k)} = \mu_i^{(k)}/t(r)$ and $t(r)$ is the base threshold defined
in Eq.~(\ref{eq:threshold}). The threshold is then obtained by a projection onto the allowed interval
\begin{equation}
t_i^{(k)} =
\bigl[\tau_i^{(k)}\bigr]_{0.3\,t(r)}^{2.0\,t(r)}
\label{eq:adaptive_threshold}
\end{equation}
where $[x]_{a}^{b} = \max(a,\min(x,b))$ denotes projection onto the interval $[a,b]$. Equation~(\ref{eq:unconstrained_threshold}) implements a proportional controller: the threshold adjustment is proportional to the deviation of $\chi_i^{(k)}$ from unity, with $A_i^{(k)}$ as the proportional gain. When $\chi_i^{(k)} > 1$, the threshold is decreased to increase detection probability; when $\chi_i^{(k)} < 1$, it is increased to reduce detections when the local discrimination signal is weaker. Equation~(\ref{eq:adaptive_threshold}) projects the result onto the interval $[0.3\,t(r),\, 2.0\,t(r)]$, preventing the threshold from departing excessively from the base value $t(r)$.

The parameter values were chosen to satisfy the following operational requirements. The learning rate $\alpha_{\mu,0}$ and its bounds $[\alpha_{\mu,\text{min}}, \alpha_{\mu,\text{max}}]$ are set so that the running mean $\mu_i^{(k)}$ converges on a timescale of $\mathcal{O}(10^2)$ measurements per setting, well within the warm-up period $K_\text{w} = 1500$, which is chosen to allow the estimates to stabilize before adaptation begins.

The reference variance $V_\text{ref}$ and the modulation bounds $[k_\text{min}, k_\text{max}]$ are chosen so that the variance-aware modulation produces a meaningful differentiation between high-variance (slow adaptation) and low-variance (fast adaptation) regimes without making the learning rate degenerate to zero or to its upper bound under typical simulation conditions.

The adaptation strength $A_0$ and the SNR thresholds $\text{SNR}_\text{low}$ and $\text{SNR}_\text{high}$ are chosen so that the threshold adjustment is proportional to the reliability of the local signal estimate: at high SNR the full adaptation strength is applied, while at low SNR the adjustment is reduced to prevent instability.

The regularization constants $\varepsilon = 10^{-6}$ and $\varepsilon' = 10^{-9}$ prevent division by zero in the variance modulation and SNR computation respectively.

The variance learning rate $\alpha_v$ is fixed and controls the time scale over which the variance estimate adapts to changes in signal variability; its value was chosen to provide a stable estimate without excessive sensitivity to single-trial fluctuations.

All parameter values were verified to produce stable adaptation behavior across the tested range of $r$; no systematic optimization was performed.

Detection probabilities for the two channels are determined using sigmoid functions with the adaptive threshold $t_i^{(k)}$
\begin{equation}
\hat{P}_{\lambda_i}^+ = \frac{1}{1 + \exp\!\bigl[-s\,(p_i - t_i^{(k)})\bigr]}
\end{equation}
\begin{equation}
\hat{P}_{\lambda_i}^- = \frac{1}{1 + \exp\!\bigl[-s\,(-p_i - t_i^{(k)})\bigr]}
\end{equation}
where $s = 20$ is the steepness parameter. The measurement outcome is then determined as in Eq.~(\ref{eq:measurement_outcome}).
Each station's threshold depends only on its local measurement history for each setting, with parameter values independent of the chosen settings. The setting dependence enters only through the net scores $p_i$ that determine the adaptation.

Table~\ref{tab:memory_results} presents the simulation results for the adaptive-threshold model with state $\ket{\Psi_r^-}$ over $r \in [-1.0, -0.1]$, with the polarizer settings $\alpha = (\SI{94.4}{\degree}, \SI{62.4}{\degree})$ and $\beta = (\SI{-6.5}{\degree}, \SI{25.5}{\degree})$ employed by Giustina et al.~\cite{Giustina2015}.

Results are obtained from $100$ independent simulations with $10^5$ pairs of entangled photons generated for each simulation. Random sampling is performed using the same Mersenne Twister generator employed in the simulations reported in Sec.~\ref{sec:results}.

For $|r| \geq 0.2$, the CH--Eberhard parameter $|J|$ decreases monotonically as $|r| \to 1$ ($J \approx -6.9\times10^{-3}$ for $r=-1$). Single detection efficiencies achieve $\eta \approx 88\%$ for $r=-1$, higher than in the base model with fixed threshold, but at the cost of a lower accuracy.

Since each station's threshold depends exclusively on its own local history and no information is shared between stations, the model is local regardless of the adaptation parameters, and $J \leq 0$ holds a priori. The relevant question is therefore not the sign of $J$ but whether the adaptive memory drives it toward the local-realistic boundary $J=0$.

Under the specific adaptive rule and parameter values tested, a paired comparison---in which the fixed- and adaptive-threshold models are initialized from the same random seed for each repeat, so that common Monte Carlo fluctuations cancel in the per-repeat difference $\Delta J = J_{\rm adaptive} - J_{\rm fixed}$ (Table~S13 in the Supplementary Material)---shows that the memory mechanism produces a systematic and reproducible shift in $J$ that never approaches the local-realistic boundary $J=0$.

The adaptive-threshold metrics of Table~\ref{tab:memory_results} are computed with the same deterministic seeding as this paired comparison, so the $J$ values reported there coincide with $J_{\rm adaptive}$; $J_{\rm fixed}$ is instead an independent Monte Carlo realization of the fixed-threshold model and is consistent with the values of Table~\ref{tab:results_psi_r} up to ordinary Monte Carlo fluctuations (within $\sim 2$ standard errors). The shift is more than ten paired standard errors from zero for $r \in \{-0.1, \ldots, -0.7\}$ ($|t|$ from $11.6$ to $50.1$), ranging from $\Delta J = -8.7\times10^{-3}$ at $r=-0.1$ to $+6.0\times10^{-3}$ at $r=-0.4$, and is not resolved above the Monte Carlo noise for $r \in \{-0.8, -0.9, -1.0\}$, where $|t|\le 2.5$ and $|\Delta J|\lesssim 5\times10^{-4}$. This result applies only to the specific rule and parameter values tested here and does not generalize to arbitrary local memory mechanisms.

\section{Conclusions}

The presented model is local by construction: detection outcomes at each station depend only on the local hidden variable $\lambda_i$ and the local measurement angle $\rho$. No information is shared between stations after the hidden variable is assigned at the source. The CH--Eberhard parameter $J < 0$ throughout the full range of entanglement parameters (Tables~\ref{tab:results_phi_r} and~\ref{tab:results_psi_r}, Figs.~\ref{fig:diagnostics_phi}b, \ref{fig:diagnostics_psi}b) is consistent with $J\leq 0$ expected a priori for a strictly local model; the substantive comparison is with the quantum benchmark $J_{\rm QM}$ (Table~\ref{tab:J_comparison_psi_r}). The same holds for the adaptive-threshold variant (Table~\ref{tab:memory_results}).

The model violates the fair-sampling assumption and exploits the detection loophole through a setting-dependent detection efficiency mechanism, approximating quantum correlations locally in the postselected detected sample.

The results are valid within the following structural boundaries. The model violates fair sampling by construction: detection probabilities depend on the local measurement setting, selecting a biased subset of the generated pairs; the scoring mechanism uses quantum-like correlation functions and coincidence probabilities evaluated locally at each station, without inter-station communication; the threshold function $t(r)$ is calibrated to a target detection efficiency via cross-validation over randomly generated polarizer configurations, ensuring generalization beyond the training set; the simulations cover specific measurement configurations and entanglement parameters; the results do not constitute a full characterization of the model over the space of all possible polarizer angles and states; the model is not a microscopic theory of photodetection; the adaptive-threshold mechanism, implementing a form of local one-sided memory, yields $J < 0$ for the specific update rule, configuration, and parameter values tested, and the adaptive-memory result applies only to the specific update rule and parameter values tested.

The CHSH violations observed in the postselected sample derive from the model's hidden-variable-dependent detection, with setting-dependent efficiency for $|r|<1$; the standard CHSH inequality does not apply because fair sampling is violated by construction. The hidden-variable distribution $\mathcal{N}(\theta, \sigma_G^2)$ with $\sigma_G$ as defined represents one specific choice; alternative distributions or parametrizations of $\sigma_G$ would yield different model behavior and remain to be explored.

The model does not provide a physical explanation of the correlations observed in loophole-free Bell experiments~\cite{Hensen2015, Giustina2015,Shalm2015}. Those experiments are conducted under conditions that specifically exclude the detection loophole exploited here. The model implements a local postselection mechanism that illustrates how setting-dependent detection efficiency shapes Bell-test statistics across a continuous range of entanglement parameters, and characterizes which structural features---threshold function, hidden-variable distribution, and local memory---affect proximity to quantum predictions.

\section*{Code Availability}

The simulation source code, the raw calibration output, and the routine used to fit the threshold function are available as Supplementary Material, together with Tables~S1--S13. The source code regenerates all tables of the main text and the Supplementary Material.

\end{document}